\documentclass[10pt,aps,prb,twocolumn,amsmath,amssymb,superscriptaddress]{revtex4-1}
\usepackage{amsmath}
\usepackage{graphicx}
\usepackage{color}
\usepackage{multirow}
\usepackage{subcaption}
\usepackage{array}
\usepackage{mhchem}
\usepackage{mathtools}
\usepackage{siunitx}
\usepackage{centernot}
\usepackage{tikz}
\tikzset{font={\fontsize{11pt}{12}\selectfont}}
\usetikzlibrary{shapes,arrows}

\makeatletter

\renewcommand*{\p@subsection}{}

\renewcommand*{\p@subsubsection}{}
\makeatother

\usepackage{hyperref}
\hypersetup{
	linkcolor=blue,          % color of internal links (change box color with linkbordercolor)
	citecolor=blue,        % color of links to bibliography
	urlcolor=blue,           % color of external links
}

\newcommand{\ket}[1]{|#1\rangle}

\newcommand{\inner}[2]{\langle#1|#2\rangle}

\newcommand{\expecth}[3]{\langle#1|#2|#3\rangle}

\begin{document}
\title{Multireference configuration interaction and perturbation theory without reduced density matrices}

\author{Ankit Mahajan}
%\email{ankitmahajan76@gmail.com}
\affiliation{Department of Chemistry, University of Colorado, Boulder, CO 80302, USA}

\author{Nick S. Blunt}
\affiliation{Department of Chemistry, Lensfield Road, Cambridge, CB2 1EW, United Kingdom}

\author{Iliya Sabzevari}
\affiliation{Department of Chemistry, University of Colorado, Boulder, CO 80302, USA}

\author{Sandeep Sharma}
\email{sanshar@gmail.com}
\affiliation{Department of Chemistry, University of Colorado, Boulder, CO 80302, USA}
\begin{abstract}
The computationally expensive evaluation and storage of high-rank reduced density matrices (RDMs) has been the bottleneck in the calculation of dynamic correlation for multireference wave functions in large active spaces. We present a stochastic formulation of multireference configuration interaction (MRCI) and perturbation theory (MRPT) that avoids the need for these expensive RDMs. The algorithm presented here is flexible enough to incorporate a wide variety of active space reference wave functions, including selected configuration interaction, matrix product states, and symmetry-projected Jastrow mean field wave functions. It enjoys the usual attractive features of Monte Carlo methods, such as embarrassing parallelizability and low memory costs. We find that the stochastic algorithm is already competitive with the deterministic algorithm for small active spaces, containing as few as 14 orbitals.
% and quickly becomes more efficient for larger active spaces.
We illustrate the utility of our stochastic formulation using benchmark applications.

\end{abstract}
\maketitle

\section{Introduction}
A quantitative treatment of electronic structure in molecules with strong electron interactions has been a challenge for quantum chemical methods. %Many of the methods developed to accomplish this are capable of accurately describing only a part of the correlation.
It is often useful to distinguish between two flavors of electron correlation: static and dynamic. Static correlation is a result of nearly degenerate electronic states and strong interactions between them. A telltale sign of this type of correlation is the dramatic failure of single-reference methods like Hartree-Fock (HF) accompanied by divergences in M\o ller-Plesset perturbation theory and coupled cluster theory. Multireference (MR) methods overcome this shortcoming by treating all (or a large number of) configurations in an active space on an equal footing. The active space usually consists of chemically relevant nearly degenerate valence orbitals. Examples of MR methods include full configuration interaction (FCI), density matrix renormalization group (DMRG),\cite{white1999ab,chan2011density} semistochastic heat bath configuration interaction (SHCI),\cite{Holmes2016b,sharma2017semistochastic} full configuration interaction quantum Monte Carlo (FCIQMC),\cite{BooThoAla-JCP-09,CleBooAla-JCP-10,PetHolChaNigUmr-PRL-12} and their self consistent field extensions, known as complete active space self consistent field (CASSCF),\cite{roos1980complete,roos1980complete2,siegbahn1981complete} DMRG-SCF,\cite{ghosh2008orbital,zgid2008density,yanai2009accelerating} SHCI-SCF,\cite{smith2017cheap} and FCIQMC-SCF,\cite{Thomas2015_3,li2016combining} respectively. While CASSCF is limited to rather small active spaces (usually less than 20 electrons and orbitals), the rest of these methods have been used in considerably larger active spaces.\cite{hachmann2007radical,marti2008density,kurashige2009high,kurashige2013entangled,sharma2014low,olivares2015ab,mussard2017one,Junhao2018,booth2013towards,li2018understanding}

Dynamic correlation is related to the fact that the usual Gaussian orbitals are inefficient at describing the Coulomb hole of an electron, which results in an extremely slow convergence of the correlation energy with the number of basis functions. To overcome this difficulty, both reasonably large basis sets (usually triple zeta basis) and explicitly correlated terms, which are functions of the inter-electronic distance, are needed to obtain chemical accuracy. The difficulty of having to use large orbital spaces is somewhat mitigated by the fact it is sufficient to use wave functions that explicitly contain only single or double excitations. MR methods suitable for capturing dynamic correlations include perturbation theory (e.g. complete active space second order perturbation (CASPT2) theory\cite{andersson1990second,andersson1992second} and n-electron valence perturbation (NEVPT) theory),\cite{angeli2001introduction,angeli2002n,angeli2004quasidegenerate} configuration interaction (e.g. multireference configuration interaction (MRCI) approaches)\cite{werner1988efficient,knowles1988efficient,knowles1992internally}, multireference coupled cluster theories (MRCC),\cite{bartlett2007coupled} canonical transformation (CT) theory,\cite{yanai2006canonical,neuscamman2010review} or the driven similarity renormalization group (DSRG) method.\cite{evangelista2014driven}

Efficient implementations of all flavors of multireference theories that are used to capture dynamic correlation utilize internally contracted states (we will describe them in more detail in Section~\ref{sec:contraction}). The great advantage of using internally contracted states is that the cost of the calculation no longer scales exponentially with the size of the active space, however, the disadvantage is that the memory cost of the calculation is a high order polynomial of the active space size. For example, in perturbation theory and configuration interaction theory, up to fourth-order reduced density matrices in the active space are needed. The cost of storing these reduced density matrices scales as the 8$^{\text{th}}$ power of the number of orbitals in the active space. Although this is still manageable for small active spaces containing 20 or fewer orbitals (the limit of a CASSCF calculation), it becomes prohibitive for the large active space calculations of the type that can be performed using modern MR methods such as DMRG, FCIQMC and selected CI. \cite{kurashige2011second,guo2016n,roemelt2016projected} Various approaches and approximations have been proposed in the past, including (a) the use of cumulant approximation,\cite{zgid2009study,saitow2013multireference,kurashige2014complete,saitow2015fully,shirai2016computational,phung2016cumulant,yanai2017multistate,nakatani-caspt2} (b) storing 4-RDM as batches of transition 3-RDM on disk,\cite{wouters2016dmrg} (c) uncontracting terms that require 4-RDMs (partial contraction),\cite{celani2000multireference,celani2000multireference,shamasundar2011new} (d) treating some terms that require 4-RDMs using matrix product states,\cite{sharma2014communication,sharma2017combining} and (e) performing time-propagation.\cite{sokolov2017time} Each of these approaches have shortcomings, for example, the cumulant approximation is highly unstable and leads to significant intruder state problems that can sometimes be fixed by including a level shift. Although the approaches (b) and (c) reduce the memory cost, they still require one to generate and store the 3-RDM which is quite expensive. The apparent exponential scaling of the uncontracted terms in (c) can be avoided using matrix product states perturbation theory.\cite{sharma2014communication,sharma2015multireference,sharma2016quasi,sharma2017combining} Finally, the time-dependent approach completely eliminates the need to store RDMs, but so far it has only been demonstrated to work with NEVPT2, where the special structure of the zeroth-order Dyall's Hamiltonian is used and it remains to be seen if this approach can be extended to more general perturbation theories and configuration interaction.

In this article, we present a stochastic formulation of the MRCI and NEVPT2 methods, that eliminates the need for constructing and storing the reduced density matrices. To reduce the number of parameters, we use a higher level of contraction, than the internal contraction, called strong contraction (SC).\cite{angeli2001introduction} We use an algorithm that is essentially identical to the Variational Monte Carlo (VMC) to optimize the wave function and calculate the energies of the SC-MRCI, and we term this approach SC-MRCI(s). This algorithm is extended to sample the Davidson size-consistency correction allowing us to calculate the SC-MRCI+Q energy up to small stochastic noise. For SC-NEVPT2(s), we again use the stochastic method to calculate the norms and energies of the perturber states, which allows us to calculate the first-order wave function and second-order energy corrections. The approach here is agnostic to the type of wave function used for performing the active space calculation and is compatible with FCI, selected CI, matrix product states, and symmetry-projected Jastrow mean field states.\cite{tahara2008variational,neuscamman2012size,neuscamman2013jastrow,mahajan2019symmetry} %We show that these methods are efficient alternatives to the deterministic formulations when larger active spaces are used. Combined with polynomially scaling active space solvers, both SC-MRCIQMC and SC-NEVPTQMC exhibit an overall polynomial scaling with system size.

The rest of this article is organized as follows: First, we briefly review the wave functions arising from the various contraction schemes and follow with the presentation of the SC-MRCI(s) and SC-NEVPT(s) algorithms. We also report details of our implementation when the FCI or selected CI methods are used to obtain the reference state. Finally, we report benchmark calculations performed using these methods, along with comparisons with the deterministic MRCI and NEVPT2 algorithms.

\section{Theory}\label{sec:theory}
\subsection{Overview of contraction schemes}\label{sec:contraction}
The zeroth-order wave function $\ket{\phi_0}$ is assumed to be an accurate representation of the exact wave function in the active space. It can be obtained by one of the several methods listed in the introduction, although in this work we will only be using CASSCF. Next, we begin by looking at various contraction schemes used in multireference theories to calculate dynamical correlation.

In the uncontracted methods, the wave function spans the entire first-order interacting space (FOIS),\cite{mclean1973classification} which consists of all singly and doubly excited determinants that couple to the active space reference through the Hamiltonian (our description is different than one in which FOIS is described as the space spanned by internally contracted states). It is only possible to use the uncontracted scheme with relatively small systems (and active spaces). Various contraction schemes have been proposed to tackle larger systems. In the fully internally contracted (FIC)\cite{Meyer1977,siegbahn1980direct} approach, single and double excitation operators are directly applied to $\ket{\phi_0}$ instead of individual determinants, resulting in a more compact wave function given by
\begin{align}
   \ket{\psi_{\text{FIC}}} =& c_0\ket{\phi_0} + \sum_{p,a}c_p^aa^{\dagger}_aa_p\ket{\phi_0} + \sum_{p,q}c_p^qa^{\dagger}_qa_p\ket{\phi_0}\nonumber\\ +\sum_{p,q,a,b}&c_{pq}^{ab}a^{\dagger}_aa^{\dagger}_ba_pa_q\ket{\phi_0}+\sum_{p,q,r,a}c_{pq}^{ra}a^{\dagger}_ra^{\dagger}_aa_pa_q\ket{\phi_0}\nonumber\\
   +\sum_{p,q,r,s}&c_{pq}^{rs}a^{\dagger}_ra^{\dagger}_sa_pa_q\ket{\phi_0},
\end{align}
where $a, b, \dots$ denote virtual orbitals, while $p, q, \dots$ denote internal (core and active) ones. Notice that the number of parameters in the FIC scheme is at most quartic in the size of the active space, unlike the uncontracted scheme, where it is exponential. Despite the enormous reduction in the number of parameters, the FIC results are usually in very good agreement with those of the uncontracted methods.
One of the drawbacks of the FIC scheme relative to the uncontracted wave function is that the internally contracted states ($a^{\dagger}_aa^{\dagger}_ba_pa_q\ket{\phi_0}$) are no longer orthogonal, which often leads to ill-conditioned generalized eigenvalue problems. In most algorithms, this difficulty is overcome by explicitly diagonalizing the overlap matrix and eliminating the zero-eigenvectors.
The IC approximation also requires construction of up to rank five RDMs which becomes very expensive as the size of the active space increases. One of the ways of avoiding this is to uncontract certain classes of excitations leading to a partially contracted (PC) scheme. This has been employed in Werner's group in the development of Werner-Knowles (WK)\cite{werner1988efficient,knowles1988efficient} and later the improved Celani-Werner (CW) MRCI\cite{celani2000multireference,shamasundar2011new}. It performs very well for smaller active spaces but has an exponential bottleneck.

The strong contraction (SC) approximation alleviates the non-orthogonality problem by further contracting the subspaces through Hamiltonian matrix elements. Here, we follow the notation introduced by Malrieu \emph{et al.}\cite{angeli2001introduction} Let $S^{(k)}_l$ denote the subspace of FIC-FOIS, where $k$ is the change in the number of active electrons $(-2\leq k\leq 2)$ and $l$ denotes the configuration of electrons in the core and virtual spaces. In the SC theory, only a single state $|\psi_l^{(k)}\rangle$ from each $S^{(k)}_l$ is used. Specifically,
\begin{align}
|\psi_{l}^{(k)}\rangle =& P_l^{(k)} H \ket{\phi_0},
\end{align}
where $P_l^{(k)}$ is the projector onto the $S_l^{(k)}$ space. Equivalently, this state is obtained by eliminating the active indices of the FIC states by contracting with the one and two electron integrals. For example, the following state from the $S^{(-2)}_{ab}$ subspace is used:
\begin{align}
|\psi_{ab}^{(-2)}\rangle=&\sum_{pq}   \left(\langle ab|pq\rangle - \langle ab|qp\rangle\right)a^{\dagger}_aa^{\dagger}_ba_pa_q\ket{\phi_0}.
\end{align}
An exception to this rule is made for the $S^{(0)}_0$ space, which is represented by the state $\ket{\phi_0}$. Implicit in this simplification is the assumption that $\ket{\phi_0}$ is the eigenstate of the active space Hamiltonian ($H_0$), otherwise an additional term $H_0\ket{\phi_0}$ should also be included. Using these states the SC wave function is described as
\begin{equation*}
   \ket{\psi_{\text{SC}}} = \sum_{k,l}c^{(k)}_l\ket{\psi^{(k)}_l}.
\end{equation*}
Note that the states $|\psi_l^{(k)}\rangle$ are mutually orthogonal (although not normalized).

Neese \emph{et al.} have argued that using the SC approximation does not lead to a large gain in efficiency in the deterministic MRCI algorithm.\cite{sivalingam2016comparison} In our stochastic formulation, we found the SC approximation to lead to a much easier optimization problem. It is also known to help avoid the intruder state problems in perturbation theory. Thus we will focus on the SC methods below. It is worth mentioning that an alternative scheme called the external contraction was proposed by Siegbahn\cite{siegbahn1980direct} which effectively eliminates the virtual orbitals from the calculation. However, it scales exponentially with the size of the active space and in its current form is only applicable when the active space wave function is expressed as a linear combination of determinants.\cite{haibo}

\subsection{SC-MRCI(s)}\label{sec:scci}
In this section, we outline the use of the SC-MRCI wave function as a VMC ansatz and its optimization using our improved orbital space VMC algorithm. In VMC, the energy of a wave function $\ket{\psi(\mathbf{p})}$, where $\mathbf{p}$ is the set of parameters, can be computed using importance sampling as
\begin{equation}
\begin{split}
 \dfrac{\langle\psi(\mathbf{p})|H|\psi(\mathbf{p})\rangle}{\inner{\psi(\mathbf{p})}{\psi(\mathbf{p})}}&= \sum_n \dfrac{|\inner{n}{\psi(\mathbf{p})}|^2}{\inner{\psi(\mathbf{p})}{\psi(\mathbf{p})}}\dfrac{\expecth{n}{H}{\psi(\mathbf{p})}}{\inner{n}{\psi(\mathbf{p})}}\\
 &= \bigg\langle\dfrac{\expecth{n}{H}{\psi(\mathbf{p})}}{\inner{n}{\psi(\mathbf{p})}}\bigg\rangle_{\rho_n},
\label{eq:vmc}
\end{split}
\end{equation}
where $\ket{n}$ is a Slater determinant walker,  $\rho_n=\dfrac{|\inner{n}{\psi(\mathbf{p})}|^2}{\inner{\psi(\mathbf{p})}{\psi(\mathbf{p})}}$ is the probability distribution used for Monte Carlo sampling. The quantity sampled is called local energy, given by
\begin{equation}
   E_L\left[n\right] = \dfrac{\expecth{n}{H}{\psi(\mathbf{p})}}{\inner{n}{\psi(\mathbf{p})}} = \sum_n\expecth{n}{H}{m}\dfrac{\inner{m}{\psi(\mathbf{p})}}{\inner{n}{\psi(\mathbf{p})}}.
\label{eq:locE}
\end{equation}
For an efficient calculation of local energy, it is essential to be able to calculate the walker overlap ratios ($\dfrac{\inner{m}{\psi(\mathbf{p})}}{\inner{n}{\psi(\mathbf{p})}}$) appearing in the above equations cheaply.

We now consider these overlaps carefully for the SC-MRCI wave function. Note that each walker belongs to a unique $S^{(k)}_l$ subspace. We will denote the subspace a walker $\ket{n}$ belongs to by $S^{(k_n)}_{l_n}$. Thus the overlap of the walker with a general contracted state is given by
\begin{equation}
    \inner{n}{\psi^{(k)}_{l}} =
		\begin{cases}
		\delta_{k,0}\delta_{l,0}\inner{n}{\phi_0}, \quad &\text{if}\ \ket{n}\in S^{(0)}_0,\\
		\delta_{k,k_n}\delta_{l,l_n}\expecth{n}{H}{\phi_0}, \quad &\text{if}\ \ket{n}\notin S^{(0)}_0.\\
		\end{cases}
\label{eq:ovlp}
\end{equation}
The overlap of the walker with $\ket{\psi_{\text{SC}}}$ is given by
\begin{equation}
   \inner{n}{\psi_{\text{SC}}} = c^{(k_n)}_{l_n}\inner{n}{\psi^{(k_n)}_{l_n}}.
\end{equation}
As $\ket{\phi_0}$ is assumed to be a selected CI wave function, its overlap with a walker can be calculated in an average $O(1)$ time by storing the determinants and coefficients in a hash table. We note that the local energy calculation is feasible for any wave function that allows efficient evaluation of these overlaps.

The walkers are sampled from the probability distribution $\rho_n$ by using the rejection free continuous time Monte Carlo (CTMC) algorithm,\cite{BORTZ197510,GILLESPIE1976403} the details of which can be seen in Ref.\citenum{Sabzevari18}. All quantities required for CTMC sampling are calculated and stored during the local energy evaluation at no additional cost.

We now turn to the problem of optimizing the wave function parameters, which in this case are the CI coefficients. At first glance, it would appear that analogous to the deterministic algorithms, the linear method\cite{nightingale2001optimization,umrigar2007alleviation,toulouse2007optimization,toulouse2008full} should be able to optimize this linearly parametrized wave function in a single step. While this is true in theory, the linear method becomes very expensive for a large number of parameters both in terms of time and memory cost.\cite{zhao2017blocked,sabzevari2019accelerated} In our experiments, we have found that it can only be feasibly applied when the number of coefficients is less than about 50,000. Since the number of states in an MRCI wave function often exceeds this number, the linear method does not appear to be a suitable choice in this problem. We instead choose AMSGrad, an accelerated stochastic gradient method developed in the machine learning community.\cite{Reddi2018} Its utility in wave function optimization in VMC has recently been reported.\cite{schwarz2017projector,Sabzevari18,mahajan2019symmetry,otis2019complementary} It should be noted that SGD has been used implicitly to solve CI problems in FCIQMC and its many variants.\cite{schwarz2017projector} AMSGrad requires an estimate of the energy gradient, which can be sampled as
\begin{equation}
   \mathbf{g}_i = \dfrac{\partial E}{\partial p_i} = \bigg\langle\dfrac{\inner{\psi_i(\mathbf{p})}{n}}{\inner{\psi(\mathbf{p})}{n}}(E_L(n)-E)\bigg\rangle_{\rho_n},
\end{equation}
where $\ket{\psi_i(\mathbf{p})} = \bigg\vert\dfrac{\partial\psi(\mathbf{p})}{\partial p_i}\bigg\rangle$ is the wave function derivative and $E$ is the energy of the wave function. For $\ket{\psi_{\text{SC}}}$, the wave function derivative overlaps are given by
\begin{equation}
	\inner{n}{\psi_{k,l}} = \delta_{k,k_n}\delta_{l,l_n}\inner{n}{\psi^{(k)}_{l}},
\end{equation}
where $\ket{\psi_{k,l}}$ denotes the derivative with respect to $c^{(k)}_l$. These quantities are needed for local energy calculation (Equation~\ref{eq:ovlp}) and are thus easily obtained.

The Davidson size-consistency correction is given by
\begin{equation}
   \Delta_DE = (1-a_0^2)(E_{\text{SC-MRCI}}-E_0^{(0)}),
\end{equation}
where $a_0$ is the coefficient of the normalized reference in the normalized SC-MRCI wave function. $a_0$ is not entirely trivial to obtain because we do not explicitly know the norm of each of the SC states. Instead, we calculate $a_0^2$ by first noting that
\begin{equation}
   a_0^2 = \dfrac{\inner{\phi_0}{\psi_{\text{opt}}}^2}{\inner{\phi_0}{\phi_0}\inner{\psi_{\text{opt}}}{\psi_{\text{opt}}}} = \left(c^{(0)}_0\right)^2 \dfrac{\inner{\phi_0}{\phi_0}}{\inner{\psi_{\text{opt}}}{\psi_{\text{opt}}}},
\end{equation}
where $\ket{\psi_{\text{opt}}}$ is the optimized SC-MRCI wave function and $c^{(0)}_0$ is the (known) coefficient of $\ket{\phi_0}$ in this wave function. This can be sampled using the CTMC algorithm as
\begin{align}
\left(c^{(0)}_0\right)^2\dfrac{\inner{\phi_0}{\phi_0}}{\inner{\psi_{\text{opt}}}{\psi_{\text{opt}}}}&= \sum_n \dfrac{|\inner{n}{\psi_{\text{opt}}}|^2}{\inner{\psi_{\text{opt}}}{\psi_{\text{opt}}}}\left(c^{(0)}_0\right)^2\dfrac{|\inner{n}{\phi_0}|^2}{|\inner{n}{\psi_{\text{opt}}}|^2} \nonumber\\
&=  \left\langle \delta_{k_n,0}\delta_{l_n,0}\right\rangle_{\rho_n}.
\end{align}

\subsection{SC-NEVPT2(s)}\label{sec:scnevpt}
The SC-NEVPT first-order correction is given by
\begin{equation}
		\ket{\psi^{(1)}} = \sum_{k,l\neq0} \dfrac{1}{E^{(0)}_0 - E^{(k)}_l}\ket{\psi^{(k)}_l},
\end{equation}
where, the perturber state energies $E^{(k)}_l$ are defined as
\begin{equation}
   E^{(k)}_l = \dfrac{\expecth{\psi^{(k)}_l}{H_D}{\psi^{(k)}_l}}{\inner{\psi^{(k)}_l}{\psi^{(k)}_l}}\label{eq:14}.
\end{equation}
And the second order SC-NEVPT energy correction is given by
\begin{equation}
		 E^{(2)} = \sum_{k,l\neq0}\dfrac{1}{E^{(0)}_0 - E^{(k)}_l}\dfrac{\inner{\psi^{(k)}_l}{\psi^{(k)}_l}}{\inner{\psi^{(0)}_0}{\psi^{(0)}_0}},\label{eq:nevpt2e2}
\end{equation}
In Eq.~\ref{eq:14}, $H_D$ is Dyall's Hamiltonian,\cite{dyall1995choice} the zeroth-order Hamiltonian employed in NEVPT. It is defined as
\begin{equation*}
  H_D = \sum_i^{\text{core}} \epsilon_ia^{\dagger}_ia_i + \sum_a^{\text{virtual}} \epsilon_aa^{\dagger}_aa_a + H_0,
\end{equation*}
where $i$ and $a$ denote the orbitals obtained by diagonalizing the core and virtual generalized Fock operators, respectively, and $\epsilon_i$ and $\epsilon_a$ the corresponding eigenenergies. $H_0$ is the full core-averaged Hamiltonian in the active space.

%To calculate the first-order wave function and the second-order energy we need two quantities, (a) the perturber energies $E^{(k)}_l$, and (b) the norm ratio appearing in Eq.~\ref{eq:nevpt2e2} for each perturber class. We discuss the calculation of each quantity separately.
%\begin{enumerate}
%\item
The energies $E^{(k)}_l$ can be obtained using Monte Carlo sampling of the numerator and denominator separately
\begin{equation}
	\begin{split}
			\dfrac{\langle\psi^{(k)}_l|H_D|\psi^{(k)}_{l}\rangle}{\inner{\psi_s}{\psi_s}} &= \sum_n \dfrac{|\inner{n}{\psi_s}|^2}{\inner{\psi_s}{\psi_s}}\dfrac{\inner{\psi^{(k)}_l}{n}}{\inner{\psi_s}{n}}\dfrac{\expecth {n}{H_D}{\psi^{(k)}_{l}}}{\inner{n}{\psi_s}},\\
			\dfrac{\inner{\psi^{(k)}_l}{\psi^{(k)}_l}}{\inner{\psi_s}{\psi_s}} &= \sum_n \dfrac{|\inner{n}{\psi_s}|^2}{\inner{\psi_s}{\psi_s}}\dfrac{|\inner{\psi^{(k)}_l}{n}|^2}{|\inner{\psi_s}{n}|^2},
	\end{split}
\label{eq:nevpt}
\end{equation}
where $\ket{\psi_s}$ wave function is used for importance sampling. In our calculations, we use the sampling wave function given by
\begin{equation}
   \ket{\psi_s} = \sum_{k,l}c^{(k)}_l\ket{\psi^{(k)}_l},
\end{equation}
where we choose the coefficients $c^{(k)}_l$ randomly, with the condition that $c^{(0)}_0$ is about an order of magnitude bigger than other coefficients. In principle, it is sufficient to choose any state $\ket{\psi_s}$ that has a non-zero overlap with all the pertuber states $\ket{\psi^{(k)}_l}$. The quantities in Eq. \ref{eq:nevpt} for all classes $S_l^{(k)}$ are sampled together using a single CTMC run similar to the one used for SC-MRCI(s) calculations. Note that the square norm ratios required for calculating the second-order energy correction (cf. Eq.~\ref{eq:nevpt2e2}) are also obtained in the same sampling run. In our experience, calculating the energy correction requires more sampling effort than calculating the coefficients in the wave function correction, likely because the variance of the quantities sampled to estimate the norms of the SC states is higher. Some perturber states have a small norm, which can be of the same order as the stochastic noise. This can cause numerical instabilities because the norms appear in the denominator of Eq.~\ref{eq:14}. To avoid this, the perturber states with a small norm are screened out. The screening does not cause significant error in the SC-NEVPT2 energies because the norm appears in the numerator of Eq.~\ref{eq:nevpt2e2}.

%\item To facilitate sampling, we recast the norm ratio appearing in Eq.~\ref{eq:nevpt2e2} as follows:
%\begin{equation*}
%   \dfrac{\inner{\psi^{(k)}_l}{\psi^{(k)}_l}}{\inner{\psi^{(0)}_0}{\psi^{(0)}_%0}} =\dfrac{\inner{\psi^{(k)}_l}{\psi^{(k)}_l}}{\inner{\psi^{(0)}_0}{\psi}},
%\end{equation*}
%where $\ket{\psi} = \ket{\psi^{(0)}_0} + \ket{\psi^{(1)}}$ is the first-order corrected wave function. The numerator and denominator can now be sampled separately as before:
%\begin{equation}
%	\begin{split}
%      \dfrac{\inner{\psi^{(k)}_l}{\psi^{(k)}_l}}{\inner{\psi}{\psi}} &= \sum_n %\dfrac{|\inner{n}{\psi}|^2}{\inner{\psi}{\psi}}\dfrac{|\inner{\psi^{(k)}_l}{n}%|^2}{|\inner{\psi}{n}|^2},\\
%			\dfrac{\inner{\psi^{(0)}_0}{\psi}}{\inner{\psi}{\psi}} &= \sum_n %\dfrac{|\inner{n}{\psi}|^2}{\inner{\psi}{\psi}}\dfrac{\inner{\psi^{(0)}_0}{n}}%{\inner{\psi}{n}}.
%	\end{split}
%\end{equation}
%Since we are sampling the first-order corrected wave function, it needs to be calculated beforehand in a separate MC run. In our experience, calculating the energy correction requires more sampling effort than calculating the coefficients in the wave function correction, likely because the variance of the quantities sampled to estimate the norms of the SC states is higher.
%\end{enumerate}
\subsection{Implementation}
We have implemented these algorithms for selected CI reference wave functions with core electrons uncorrelated. Consider the expression for local energy given in Eq. \ref{eq:locE}:
\begin{equation*}
   E_L\left[n\right] =  \sum_n\expecth{n}{H}{m}\dfrac{\inner{m}{\psi(\mathbf{p})}}{\inner{n}{\psi(\mathbf{p})}}.
\end{equation*}
The walker $\ket{n}$ belongs to either the CAS or the FOIS. The determinants $\ket{m}$ are generated from $\ket{n}$ through the Hamiltonian. Note that only those $\ket{m}$'s that are at most doubly excited from the CAS have a non-zero overlap with the wave function and thus only these excitations need to be generated. If the resulting $\ket{m}$ is not in the CAS, we again need to generate excitations from this determinant using the Hamiltonian to calculate its overlap with the wave function (cf. Eq. \ref{eq:ovlp}). The number of determinants that need to be generated from $\ket{m}$ is significantly less than the number of all determinants connected to it since they have to be in the CAS to have a non-zero overlap with $\ket{\phi_0}$. We use the heat-bath algorithm\cite{Holmes2016b} to generate all excitations efficiently. The determinants in the reference $\ket{\phi_0}$ are stored in a hash table, so the overlap of a determinant with it can be calculated in constant time on average. Because this method avoids calculation of the expensive RDMs, its memory cost is negligible compared to the deterministic algorithm. Similar considerations apply to the quantities sampled in SC-NEVPT2(s), with the exception that in this case, the first set of excitations are generated using Dyall's Hamiltonian instead of the full system Hamiltonian. This implementation can be extended to correlate core electrons and to work with other kinds of reference wave functions. Details of how this can be accomplished will be presented in a forthcoming publication.

\section{Results}
In this section, we will present applications of SC-MRCI(s) and SC-NEVPT2(s) with a selected CI reference to demonstrate its utility in treating multireference problems quantitatively. We compare our energies to Celani-Werner (CW) MRCI energies obtained using MOLPRO,\cite{werner2012molpro} version 2019.1. We used PySCF\cite{sun2018pyscf} to generate Hamiltonian integrals. The selected CI program Dice\cite{Holmes2016b,sharma2017semistochastic,smith2017cheap} was used to obtain the determinants in the CASSCF wave function. %In all calculations, we retained enough determinants in the selected CI wave function to get an energy within 0.2 mE\textsubscript{h} of the CASSCF energy.

We report the computational timings for SC-MRCI(s) calculations of hydrogen chains of increasing length and compare them with CW-MRCI. We then present potential energy curves and spectroscopic constants for a few diatomic molecules. Finally, we analyze the efficacy of the Davidson correction for achieving approximate size-consistency.

\subsection{Hydrogen chain}
\begin{figure}
\centering
\includegraphics[width=0.5\textwidth]{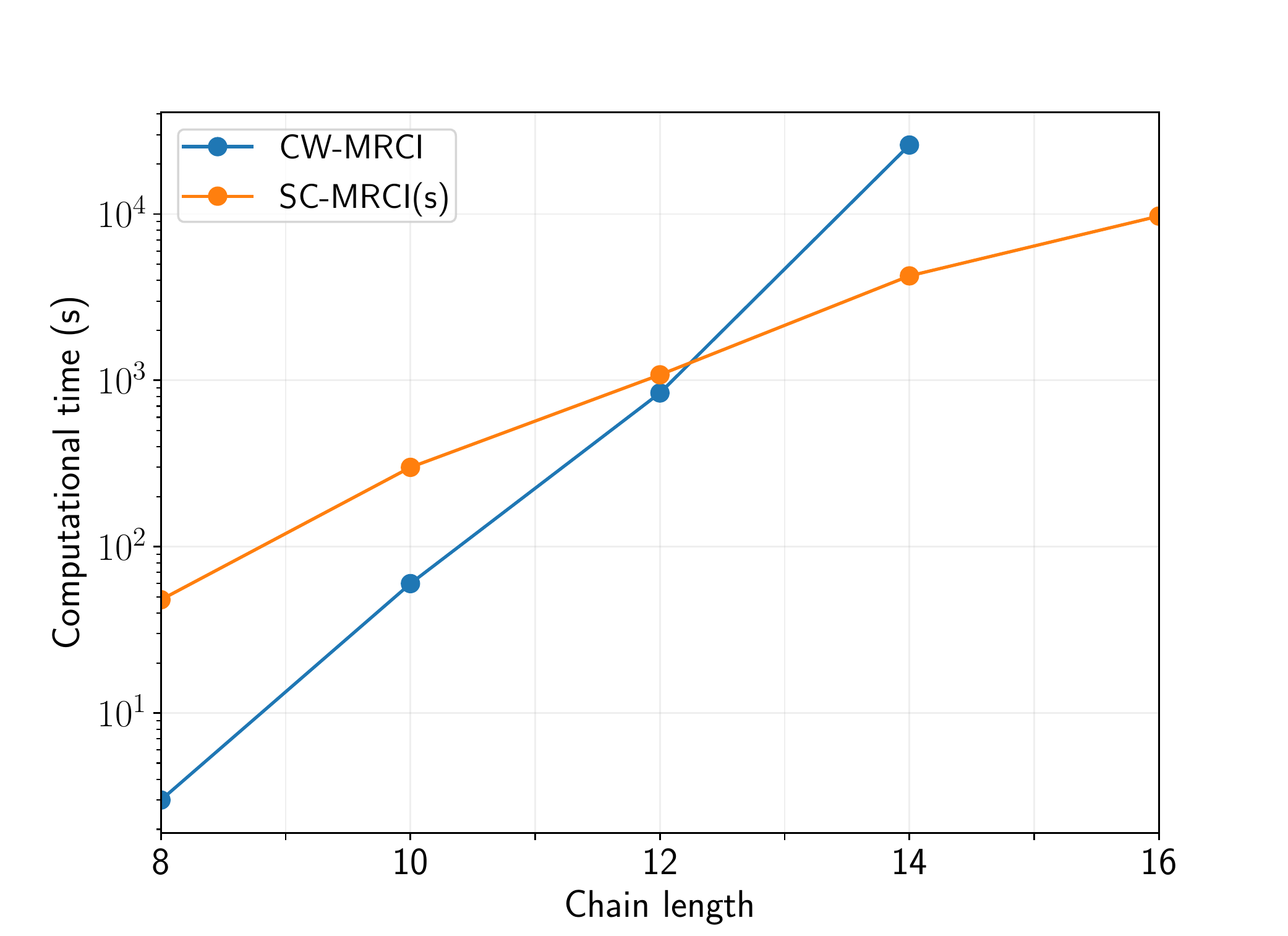}
\caption{Computational time in seconds for CW-MRCI and SC-MRCI(s) calculations of hydrogen chains H$_n$ in the 6-31g basis set using the ($n$e, $n$o) active space consisting of all 1s orbitals. The CW-MRCI calculations were performed serially while the SC-MRCI(s) calculations were performed using 24 cores.}\label{fig:h_chain}
\end{figure}
MRCI energies were calculated for open hydrogen chains of increasing lengths in the 6-31g basis set. The bond length was set to 2 Bohr in all calculations. The active space consisting of the 1s orbitals on each hydrogen was used. Fig.~\ref{fig:h_chain} presents computational times, and Table \ref{tab:h_chain} shows the ground state energies for SC-MRCI(s) and CW-MRCI.

\begin{table}[htp]
\caption{Ground state energies (E\textsubscript{h}) for hydrogen chains H\textsubscript{n} in the 6-31g basis set. Stochastic errors in the QMC calculations are less than 1 mE\textsubscript{h}.}\label{tab:h_chain}
\centering
\begin{tabular}{ccccc}
\hline
\hline
Chain length &~~~& CW-MRCI &~~& SC-MRCI(s)\\
\hline
8 && -4.416 && -4.415\\
10 && -5.518 && -5.516\\
12 && -6.620 && -6.618\\
14 && -7.722 && -7.719\\
16 && - && -8.820 \\
\hline
\end{tabular}
\end{table}

AMSGrad was used to optimize the variational energy and to obtain the converged SC-MRCI(s) wave function. It usually requires fewer AMSGrad iterations to converge to the final result if more stochastic samples are used. In our numerical experiments, we have found that using enough samples to obtain energies with an error of $\sim 10$ mE\textsubscript{h} in the first iteration leads to smooth optimization in most cases. 2400 stochastic samples were enough to achieve this accuracy for all chain lengths considered here. We progressively increased the number of stochastic samples as the wave function approached convergence, and all the reported SC-MRCI(s) energies here have a stochastic error of less than 1 mE\textsubscript{h}. Calculations were performed on a single compute node with two Intel\textsuperscript{\textregistered} Xeon\textsuperscript{\textregistered} E5-2680 v3 processors (2.5 GHz) and 116 GB memory. SC-MRCI(s) calculations were parallelized over all available cores using MPI, while MOLPRO calculations were performed serially since a parallel implementation is not available. It is apparent from Fig.~\ref{fig:h_chain} that although CW-MRCI is very efficient for smaller active spaces, its scaling with the size of the active space is worse compared to SC-MRCI(s). This can be attributed to the use of uncontracted excitations for certain semi-internal excitation classes in CW-MRCI. As a result, for even the moderately sized \ce{H14} chain, SC-MRCI(s) achieves a performance similar to CW-MRCI. For the \ce{H16} chain, we were unable to perform the CW-MRCI calculation on a single node, while the SC-MRCI(s) calculation took only about three hours. We note that CW-MRCI can be parallelized to obtain better performance but not as effectively as VMC, which is embarrassingly parallel. The timings for SC-MRCI(s) can be reduced linearly by using more processors thus offsetting the larger prefactor often present in the scaling of VMC methods. For this small basis set, the absolute differences in energies between the two methods are relatively small, with the SC approximation leading to errors of less than 3 mE\textsubscript{h}. Based on this evidence, we expect that SC-MRCI(s) can be scaled to larger active spaces by using reference wave functions such as symmetry projected Jastrow mean-field and matrix product states. With improvements to our implementation, the stochastic algorithm might compare favorably to the deterministic methods for even small to moderate-sized active spaces.

\begin{figure*}[htp]
\centering
    \begin{subfigure}[t]{0.3\textwidth}
        \centering
        \includegraphics[width=1.1\textwidth]{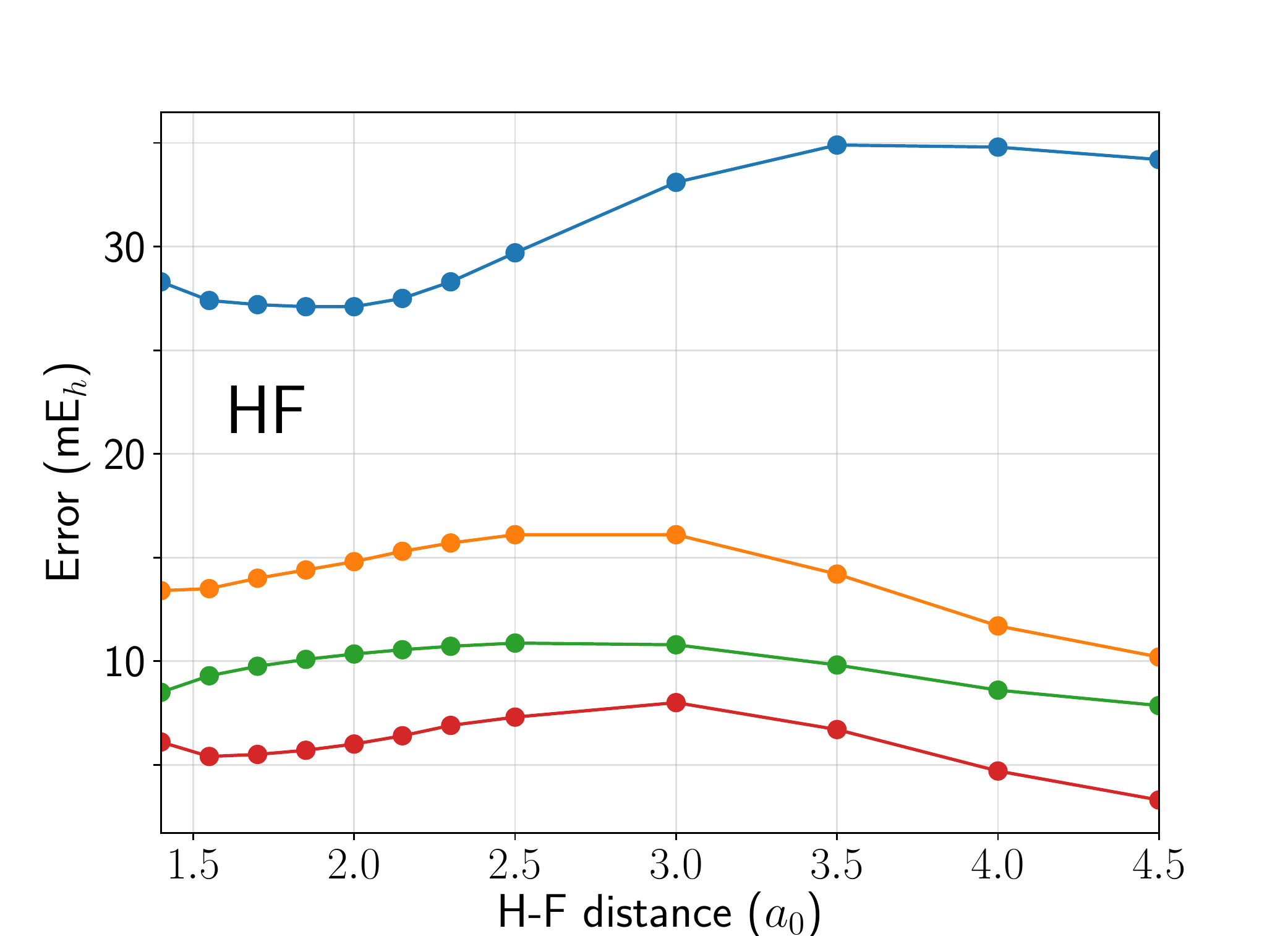}
        %\caption{HF}
    \end{subfigure}%
    ~
    \begin{subfigure}[t]{0.3\textwidth}
        \centering
        \includegraphics[width=1.1\textwidth]{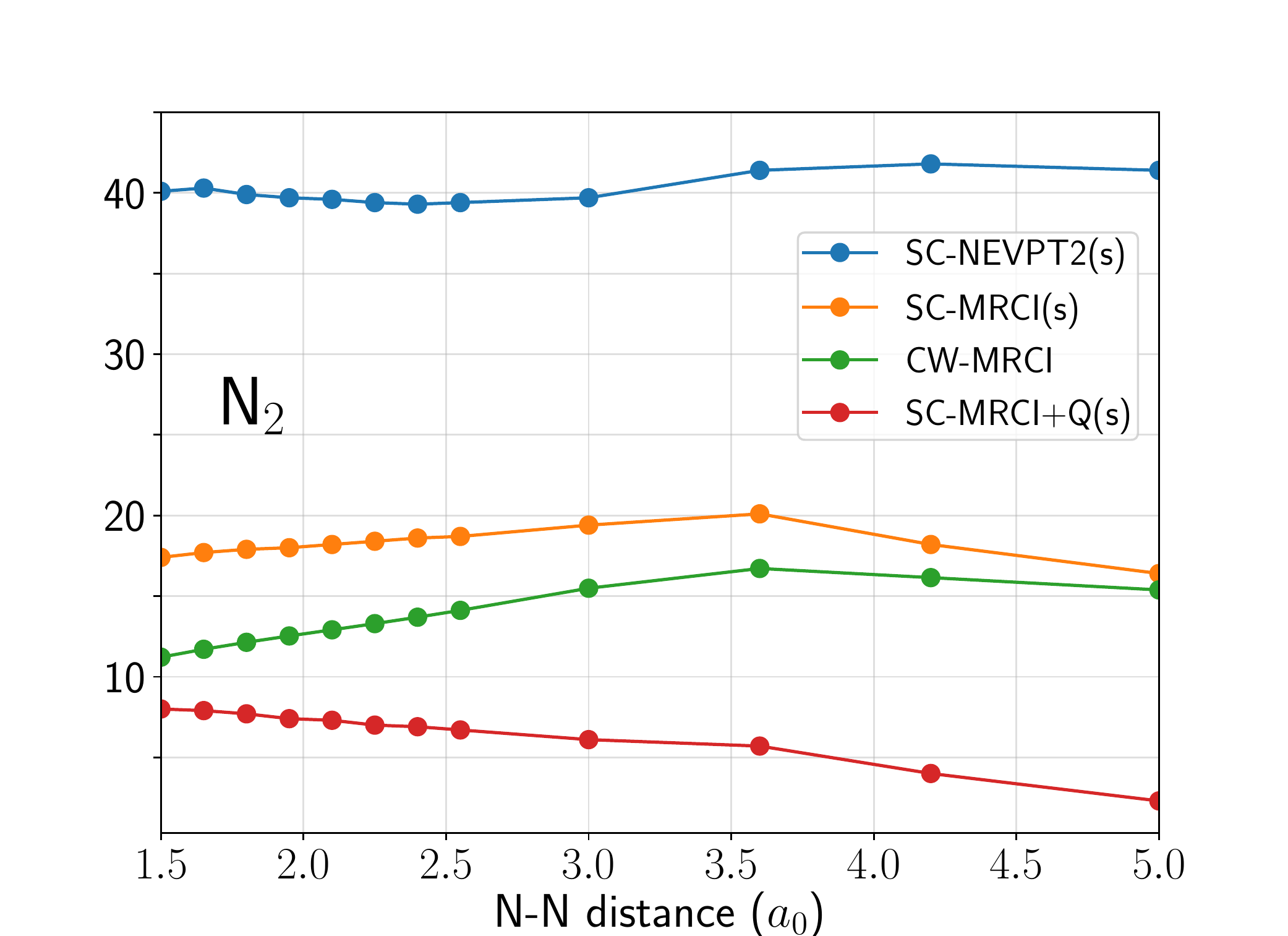}
        %\caption{\ce{N2}}
    \end{subfigure}
    ~
    \begin{subfigure}[t]{0.3\textwidth}
        \centering
        \includegraphics[width=1.1\textwidth]{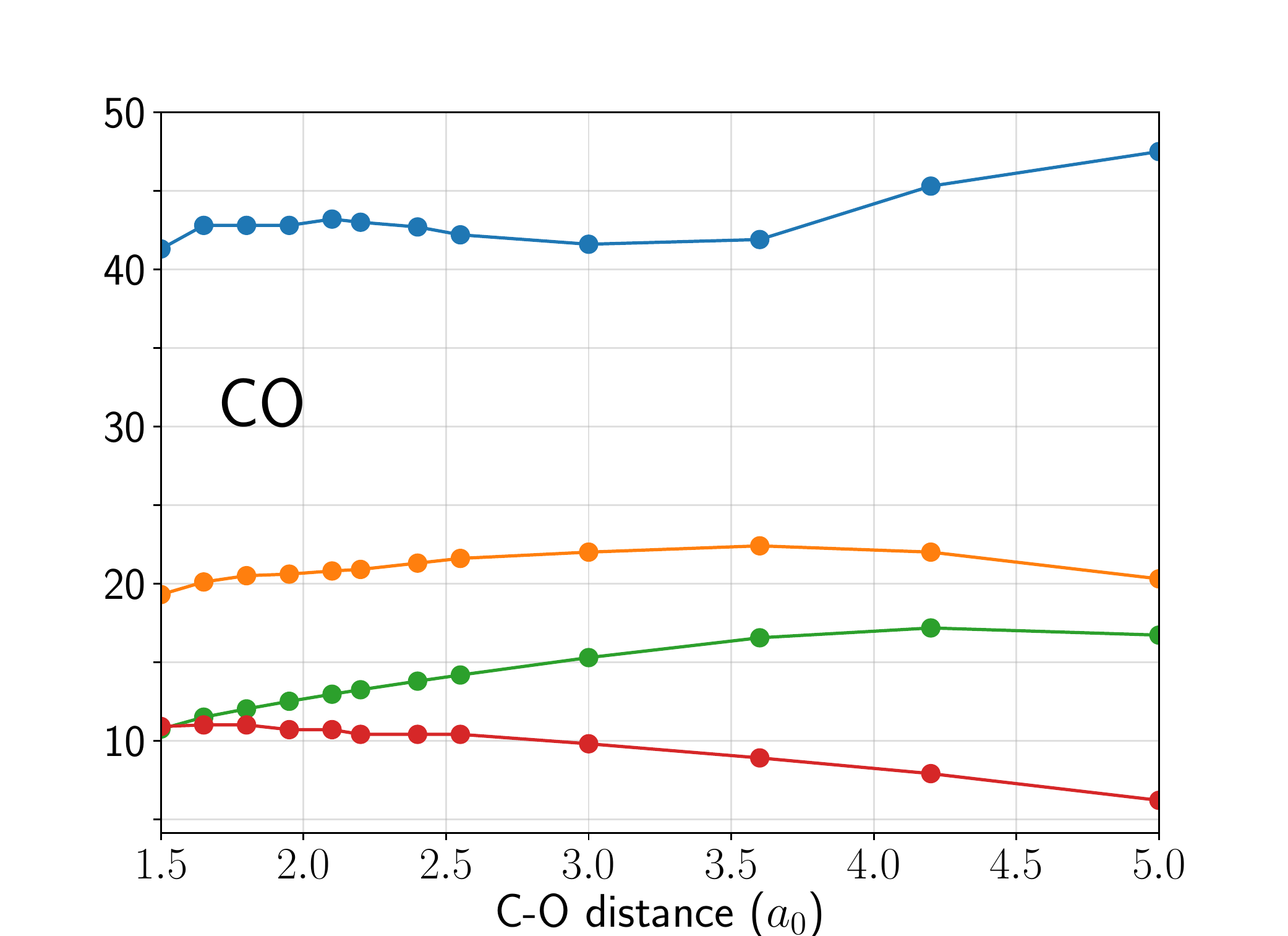}
        %\caption{CO}
    \end{subfigure}
\caption{Errors in ground state energies relative to CW-MRCI+Q for HF, \ce{N2}, and CO in the cc-pVTZ basis set. The stochastic errors in SC-MRCI(s)and SC-MRCI+Q(s) energies are smaller than the symbols. Refer to the text for a discussion on stochastic errors in the SC-NEVPT2(s) energies.}\label{fig:pes}
\end{figure*}

\subsection{Diatomic molecules}
We have calculated the ground state potential energy curves of HF, \ce{N2}, and CO molecules in the cc-pVTZ basis using full-valence active spaces of sizes (8e,5o), (10e,8o), and (10e,8o), respectively. The 1s core electrons were frozen in all correlated calculations. We used 48,000 stochastic samples in all SC-MRCI(s) calculations for each AMSGrad iteration. In all cases, the optimization converged in about 15-20 AMSgrad iterations. We then performed a final single point calculation with sufficient stochastic samples to achieve an error less than 0.2 mE\textsubscript{h}. We have found that the number of stochastic samples needed to obtain a given accuracy scales roughly linearly with the number of variables in the wave function. For SC-NEVPT2(s) calculations, $9.6\times10^6$ stochastic samples were used. The greater number of stochastic samples was needed partly because the second-order correction to the energy is a non-linear function of $E_l^{(k)}$ (see Eq.~\ref{eq:nevpt2e2}), which is itself calculated stochastically. As a result, in addition to a stochastic error, the SC-NEVPT2(s) energy also has a systematic bias, as has been investigated in other QMC methods previously.\cite{Zhao2016,blunt2018nonlinear} However, with the relatively large number of stochastic samples used here, we find that the systematic bias is smaller than the stochastic noise. We confirmed this by performing 50 independent energy calculations at one geometry for each molecule. The standard deviation of the resulting distribution of energies was less than 0.3 mE\textsubscript{h} and the average was found to be within one standard deviation of the deterministic SC-NEVPT2 energy.

\begin{table}[htp]
\centering
\caption{Non-parallelity errors (mE\textsubscript{h}) in Fig.~\ref{fig:pes}.}\label{tab:np}
\begin{tabular}{ccccccc}
\hline
\hline
Method &~~~& HF &~~& \ce{N2} &~~& CO\\
\hline
SC-NEVPT2(s) && 7.8 && 2.5 && 4.1\\
SC-MRCI(s) && 6.0 && 3.7 && 3.0\\
CW-MRCI && 3.0 && 5.5 && 6.4\\
SC-MRCI+Q(s) && 4.6 && 5.7 && 4.8\\
\hline
\end{tabular}
\end{table}

The errors in ground state energies relative to CW-MRCI+Q for the three molecules are shown in Fig.~\ref{fig:pes}. The corresponding non-parallelity (NP) errors are shown in Table~\ref{tab:np}. Out of all the methods studied here, SC-MRCI+Q(s) recovers the largest percentage of correlation energy in almost all cases. HF has a strongly avoided crossing around 3 Bohr, and all CI methods have a peak in their energy errors around this distance. SC-NEVPT2(s) has a large NP error of 8.4 mE\textsubscript{h}, while CW-MRCI and SC-MRCI+Q(s) have relatively small NP errors. For \ce{N2} and CO, SC-NEVPT2(s) is remarkably accurate with NP errors of 2.7 and 4.1 mE\textsubscript{h}, respectively. For these two molecules, SC-MRCI(s) has considerably smaller NP errors than SC-MRCI+Q(s). It should be noted that the biggest contribution to the NP error in the SC-MRCI+Q(s) method results from the most stretched geometry in all cases, where it over-correlates the system compared to near-equilibrium geometries. The SC approximation leads to larger absolute errors in MRCI compared to SC-NEVPT2. Nonetheless, the NP errors in both theories due to the SC approximation are relatively small, especially considering the compactness of the SC wave functions. We note that SC-MRCI results of similar quality have also been reported in Refs.\citenum{sivalingam2016comparison} and \citenum{angeli2012comparison}.

\begin{table}[htp]
\centering
\caption{Spectroscopic constants calculated using a degree seven polynomial fit.}\label{tab:spec}
\begin{tabular}{cclcccccc}
\hline
\hline
Molecule && Method &~~& $r_e$ ($a_0$)  &~~& $\omega_e$ (cm\textsuperscript{-1}) &~~& $\omega_ex_e$ (cm\textsuperscript{-1})\\
\hline
HF && SC-NEVPT2 && 1.747 && 4203 && 77\\
   && SC-MRCI+Q(s) && 1.743 && 4212 && 80\\
   && CW-MRCI+Q && 1.744 && 4182 && 83\\
\hline
\ce{N2} && SC-NEVPT2 && 2.090 && 2327 && 14\\
   && SC-MRCI+Q(s) && 2.090 && 2320 && 14\\
   && CW-MRCI+Q && 2.089 && 2330 && 14\\
\hline
\ce{CO} && SC-NEVPT2 && 2.151 && 2135 && 13\\
   && SC-MRCI+Q(s) && 2.149 && 2154 && 19\\
   && CW-MRCI+Q && 2.150 && 2141 && 13\\
\hline
\end{tabular}
\end{table}

Table \ref{tab:spec} shows the spectroscopic constants $r_e$, $\omega_e$, and $\omega_ex_e$ for these diatomic molecules obtained using MOLPRO. These are calculated by fitting the potential energy curve with a high-degree polynomial. To estimate the errors in the constants for SC-MRCI+Q(s), we performed a Monte Carlo analysis by fitting to energy values obtained by adding normally distributed noise, with a standard deviation chosen to match the error estimate on each data point. Sufficient samples were used so that the average values reported in the table are converged to all the digits shown. Because we do not have precise error estimates for the SC-NEVPT2(s) energies here, we opted to use deterministic SC-NEVPT2 values for these calculations. The values of constants are in excellent agreement with each other. The NP errors do not appear to affect these constants much.

Table \ref{tab:size_con} shows the size-consistency errors for these methods in the nitrogen molecule. In this case, all stochastic CI energies were converged to an accuracy of 0.2 mE\textsubscript{h}. SC-NEVPT2(s) is exactly size consistent (within statistical error) as expected. CW-MRCI and SC-MRCI(s) have large size-consistency errors which arise due to the same reasons as in the single-reference CI case. The Davidson correction does seem to remedy this issue to a large extent. The relatively small NP and size-consistency errors in the SC-MRCI+Q(s) and SC-NEVPT2(s) methods are encouraging and indicate that it may be feasible to treat problems with a large number of virtual orbitals using our framework.

\begin{table}[htp]
\centering
\caption{Size-consistency errors for \ce{N2} calculated as the difference between the energy of two well separated atoms and twice the energy of a single atom.}\label{tab:size_con}
\begin{tabular}{ccc}
\hline
\hline
Method &~~~& $\Delta E$ (mE\textsubscript{h})\\
\hline
SC-NEVPT2(s) && 0.0\\
CW-MRCI && 7.6\\
CW-MRCI+Q && 1.0\\
SC-MRCI(s) && 8.0\\
SC-MRCI+Q(s) && 2.1\\
\hline
\end{tabular}
\end{table}

\section{Conclusions}
In this work, we have presented stochastic formulations of SC-MRCI and SC-NEVPT2 methods that avoid the requirement of storing expensive high-order active space RDMs. The need to calculate and store these RDMs has been a major bottleneck in calculating dynamic correlation in multireference theories, especially when active spaces are large. Using benchmark calculations on hydrogen chains and diatomic molecules, we have argued that the stochastic method, presented here, represents an attractive alternative to the deterministic methods. It outperforms the corresponding deterministic method even with active spaces that are as small as 14 orbitals.

Our work also highlights the accuracy of the strongly contracted wave functions, which is in agreement with previous work. However, even with this relatively small loss of accuracy, the use of strong contraction in deterministic algorithms is often not recommended because the saving in computational time relative to internal or partial contraction is negligible. This metric drastically changes in the stochastic approach presented here, because the optimization problem one needs to solve for strong contraction is significantly easier than when partial or internal contraction are used. Thus with stochastic methods, our recommendation is to use strong contraction and only resort to internal contraction when it is known that the former is likely to fail.

This work will be extended in several directions. All calculations presented here have used the frozen core approximation. We are working on an efficient implementation of excitation classes that correlate core electrons. It will be interesting to see how the stochastic method performs with other reference wave functions, such as matrix product states and symmetry-projected Jastrow mean field states. A problem that we have not discussed here is that the strongly contracted wave functions are not invariant to unitary transformations in the virtual orbitals. Our preliminary results indicate that the results are most accurate when canonical CASSCF virtual orbitals are used, however, the efficiency of calculation suffers when such delocalized orbitals are used. More work is needed to determine what kind of virtual (and core) orbitals will lead to the best results. The fact that our formulation can be used with many different active-space wave functions and places virtually no restrictions on the size of the active or virtual spaces, raises the exciting prospect of performing multireference calculations on large systems that are beyond the reach of current methods.

\section{Acknowledgements}
The funding for this project was provided by the national science foundation through the grant CHE-1800584. SS was also partly supported through the Sloan research fellowship. NSB is grateful to St John's College, Cambridge for funding through a Research Fellowship.

%\bibliographystyle{achemso}
%\bibliography{ref}

\begin{mcitethebibliography}{84}
\providecommand*\natexlab[1]{#1}
\providecommand*\mciteSetBstSublistMode[1]{}
\providecommand*\mciteSetBstMaxWidthForm[2]{}
\providecommand*\mciteBstWouldAddEndPuncttrue
  {\def\EndOfBibitem{\unskip.}}
\providecommand*\mciteBstWouldAddEndPunctfalse
  {\let\EndOfBibitem\relax}
\providecommand*\mciteSetBstMidEndSepPunct[3]{}
\providecommand*\mciteSetBstSublistLabelBeginEnd[3]{}
\providecommand*\EndOfBibitem{}
\mciteSetBstSublistMode{f}
\mciteSetBstMaxWidthForm{subitem}{(\alph{mcitesubitemcount})}
\mciteSetBstSublistLabelBeginEnd
  {\mcitemaxwidthsubitemform\space}
  {\relax}
  {\relax}

\bibitem[White and Martin(1999)White, and Martin]{white1999ab}
White,~S.~R.; Martin,~R.~L. Ab initio quantum chemistry using the density
  matrix renormalization group. \emph{The Journal of chemical physics}
  \textbf{1999}, \emph{110}, 4127--4130\relax
\mciteBstWouldAddEndPuncttrue
\mciteSetBstMidEndSepPunct{\mcitedefaultmidpunct}
{\mcitedefaultendpunct}{\mcitedefaultseppunct}\relax
\EndOfBibitem
\bibitem[Chan and Sharma(2011)Chan, and Sharma]{chan2011density}
Chan,~G. K.-L.; Sharma,~S. The density matrix renormalization group in quantum
  chemistry. \emph{Annual review of physical chemistry} \textbf{2011},
  \emph{62}, 465--481\relax
\mciteBstWouldAddEndPuncttrue
\mciteSetBstMidEndSepPunct{\mcitedefaultmidpunct}
{\mcitedefaultendpunct}{\mcitedefaultseppunct}\relax
\EndOfBibitem
\bibitem[Holmes \latin{et~al.}(2016)Holmes, Tubman, and Umrigar]{Holmes2016b}
Holmes,~A.~A.; Tubman,~N.~M.; Umrigar,~C.~J. Heat-Bath Configuration
  Interaction: An Efficient Selected Configuration Interaction Algorithm
  Inspired by Heat-Bath Sampling. \emph{J. Chem. Theory Comput.} \textbf{2016},
  \emph{12}, 3674--3680, PMID: 27428771\relax
\mciteBstWouldAddEndPuncttrue
\mciteSetBstMidEndSepPunct{\mcitedefaultmidpunct}
{\mcitedefaultendpunct}{\mcitedefaultseppunct}\relax
\EndOfBibitem
\bibitem[Sharma \latin{et~al.}(2017)Sharma, Holmes, Jeanmairet, Alavi, and
  Umrigar]{sharma2017semistochastic}
Sharma,~S.; Holmes,~A.~A.; Jeanmairet,~G.; Alavi,~A.; Umrigar,~C.~J.
  Semistochastic Heat-bath Configuration Interaction method: selected
  configuration interaction with semistochastic perturbation theory. \emph{J.
  Chem. Theory Comput.} \textbf{2017}, \emph{13}, 1595--1604\relax
\mciteBstWouldAddEndPuncttrue
\mciteSetBstMidEndSepPunct{\mcitedefaultmidpunct}
{\mcitedefaultendpunct}{\mcitedefaultseppunct}\relax
\EndOfBibitem
\bibitem[Booth \latin{et~al.}(2009)Booth, Thom, and Alavi]{BooThoAla-JCP-09}
Booth,~G.~H.; Thom,~A. J.~W.; Alavi,~A. Fermion Monte Carlo without fixed
  nodes: A game of life, death, and annihilation in Slater determinant space.
  \emph{J. Chem. Phys.} \textbf{2009}, \emph{131}, 054106\relax
\mciteBstWouldAddEndPuncttrue
\mciteSetBstMidEndSepPunct{\mcitedefaultmidpunct}
{\mcitedefaultendpunct}{\mcitedefaultseppunct}\relax
\EndOfBibitem
\bibitem[Cleland \latin{et~al.}(2010)Cleland, Booth, and
  Alavi]{CleBooAla-JCP-10}
Cleland,~D.; Booth,~G.~H.; Alavi,~A. Communications: Survival of the fittest:
  Accelerating convergence in full configuration-interaction quantum Monte
  Carlo. \emph{J. Chem. Phys.} \textbf{2010}, \emph{132}, 041103\relax
\mciteBstWouldAddEndPuncttrue
\mciteSetBstMidEndSepPunct{\mcitedefaultmidpunct}
{\mcitedefaultendpunct}{\mcitedefaultseppunct}\relax
\EndOfBibitem
\bibitem[Petruzielo \latin{et~al.}(2012)Petruzielo, Holmes, Changlani,
  Nightingale, and Umrigar]{PetHolChaNigUmr-PRL-12}
Petruzielo,~F.~R.; Holmes,~A.~A.; Changlani,~H.~J.; Nightingale,~M.~P.;
  Umrigar,~C.~J. Semistochastic Projector Monte Carlo Method. \emph{Phys. Rev.
  Lett.} \textbf{2012}, \emph{109}, 230201\relax
\mciteBstWouldAddEndPuncttrue
\mciteSetBstMidEndSepPunct{\mcitedefaultmidpunct}
{\mcitedefaultendpunct}{\mcitedefaultseppunct}\relax
\EndOfBibitem
\bibitem[Roos \latin{et~al.}(1980)Roos, Taylor, Si, \latin{et~al.}
  others]{roos1980complete}
Roos,~B.~O.; Taylor,~P.~R.; Si,~P.~E., \latin{et~al.}  A complete active space
  SCF method (CASSCF) using a density matrix formulated super-CI approach.
  \emph{Chemical Physics} \textbf{1980}, \emph{48}, 157--173\relax
\mciteBstWouldAddEndPuncttrue
\mciteSetBstMidEndSepPunct{\mcitedefaultmidpunct}
{\mcitedefaultendpunct}{\mcitedefaultseppunct}\relax
\EndOfBibitem
\bibitem[Roos(1980)]{roos1980complete2}
Roos,~B.~O. The complete active space SCF method in a fock-matrix-based
  super-CI formulation. \emph{International Journal of Quantum Chemistry}
  \textbf{1980}, \emph{18}, 175--189\relax
\mciteBstWouldAddEndPuncttrue
\mciteSetBstMidEndSepPunct{\mcitedefaultmidpunct}
{\mcitedefaultendpunct}{\mcitedefaultseppunct}\relax
\EndOfBibitem
\bibitem[Siegbahn \latin{et~al.}(1981)Siegbahn, Alml{\"o}f, Heiberg, and
  Roos]{siegbahn1981complete}
Siegbahn,~P.~E.; Alml{\"o}f,~J.; Heiberg,~A.; Roos,~B.~O. The complete active
  space SCF (CASSCF) method in a Newton--Raphson formulation with application
  to the HNO molecule. \emph{The Journal of Chemical Physics} \textbf{1981},
  \emph{74}, 2384--2396\relax
\mciteBstWouldAddEndPuncttrue
\mciteSetBstMidEndSepPunct{\mcitedefaultmidpunct}
{\mcitedefaultendpunct}{\mcitedefaultseppunct}\relax
\EndOfBibitem
\bibitem[Ghosh \latin{et~al.}(2008)Ghosh, Hachmann, Yanai, and
  Chan]{ghosh2008orbital}
Ghosh,~D.; Hachmann,~J.; Yanai,~T.; Chan,~G. K.-L. Orbital optimization in the
  density matrix renormalization group, with applications to polyenes and
  $\beta$-carotene. \emph{The Journal of chemical physics} \textbf{2008},
  \emph{128}, 144117\relax
\mciteBstWouldAddEndPuncttrue
\mciteSetBstMidEndSepPunct{\mcitedefaultmidpunct}
{\mcitedefaultendpunct}{\mcitedefaultseppunct}\relax
\EndOfBibitem
\bibitem[Zgid and Nooijen(2008)Zgid, and Nooijen]{zgid2008density}
Zgid,~D.; Nooijen,~M. The density matrix renormalization group self-consistent
  field method: Orbital optimization with the density matrix renormalization
  group method in the active space. \emph{The Journal of chemical physics}
  \textbf{2008}, \emph{128}, 144116\relax
\mciteBstWouldAddEndPuncttrue
\mciteSetBstMidEndSepPunct{\mcitedefaultmidpunct}
{\mcitedefaultendpunct}{\mcitedefaultseppunct}\relax
\EndOfBibitem
\bibitem[Yanai \latin{et~al.}(2009)Yanai, Kurashige, Ghosh, and
  Chan]{yanai2009accelerating}
Yanai,~T.; Kurashige,~Y.; Ghosh,~D.; Chan,~G. K.-L. Accelerating convergence in
  iterative solution for large-scale complete active space
  self-consistent-field calculations. \emph{International Journal of Quantum
  Chemistry} \textbf{2009}, \emph{109}, 2178--2190\relax
\mciteBstWouldAddEndPuncttrue
\mciteSetBstMidEndSepPunct{\mcitedefaultmidpunct}
{\mcitedefaultendpunct}{\mcitedefaultseppunct}\relax
\EndOfBibitem
\bibitem[Smith \latin{et~al.}(2017)Smith, Mussard, Holmes, and
  Sharma]{smith2017cheap}
Smith,~J.~E.; Mussard,~B.; Holmes,~A.~A.; Sharma,~S. Cheap and near exact
  CASSCF with large active spaces. \emph{J. Chem. Theory Comput.}
  \textbf{2017}, \emph{13}, 5468--5478\relax
\mciteBstWouldAddEndPuncttrue
\mciteSetBstMidEndSepPunct{\mcitedefaultmidpunct}
{\mcitedefaultendpunct}{\mcitedefaultseppunct}\relax
\EndOfBibitem
\bibitem[Thomas \latin{et~al.}(2015)Thomas, Sun, Alavi, and
  Booth]{Thomas2015_3}
Thomas,~R.~E.; Sun,~Q.; Alavi,~A.; Booth,~G.~H. Stochastic Multiconfigurational
  Self-Consistent Field Theory. \emph{J. Chem. Theory Comput.} \textbf{2015},
  \emph{11}, 5316\relax
\mciteBstWouldAddEndPuncttrue
\mciteSetBstMidEndSepPunct{\mcitedefaultmidpunct}
{\mcitedefaultendpunct}{\mcitedefaultseppunct}\relax
\EndOfBibitem
\bibitem[Li~Manni \latin{et~al.}(2016)Li~Manni, Smart, and
  Alavi]{li2016combining}
Li~Manni,~G.; Smart,~S.~D.; Alavi,~A. Combining the complete active space
  self-consistent field method and the full configuration interaction quantum
  Monte Carlo within a super-CI framework, with application to challenging
  metal-porphyrins. \emph{Journal of chemical theory and computation}
  \textbf{2016}, \emph{12}, 1245--1258\relax
\mciteBstWouldAddEndPuncttrue
\mciteSetBstMidEndSepPunct{\mcitedefaultmidpunct}
{\mcitedefaultendpunct}{\mcitedefaultseppunct}\relax
\EndOfBibitem
\bibitem[Hachmann \latin{et~al.}(2007)Hachmann, Dorando, Avil{\'e}s, and
  Chan]{hachmann2007radical}
Hachmann,~J.; Dorando,~J.~J.; Avil{\'e}s,~M.; Chan,~G. K.-L. The radical
  character of the acenes: A density matrix renormalization group study.
  \emph{The Journal of chemical physics} \textbf{2007}, \emph{127},
  134309\relax
\mciteBstWouldAddEndPuncttrue
\mciteSetBstMidEndSepPunct{\mcitedefaultmidpunct}
{\mcitedefaultendpunct}{\mcitedefaultseppunct}\relax
\EndOfBibitem
\bibitem[Marti \latin{et~al.}(2008)Marti, Ond{\'\i}k, Moritz, and
  Reiher]{marti2008density}
Marti,~K.~H.; Ond{\'\i}k,~I.~M.; Moritz,~G.; Reiher,~M. Density matrix
  renormalization group calculations on relative energies of transition metal
  complexes and clusters. \emph{The Journal of chemical physics} \textbf{2008},
  \emph{128}, 014104\relax
\mciteBstWouldAddEndPuncttrue
\mciteSetBstMidEndSepPunct{\mcitedefaultmidpunct}
{\mcitedefaultendpunct}{\mcitedefaultseppunct}\relax
\EndOfBibitem
\bibitem[Kurashige and Yanai(2009)Kurashige, and Yanai]{kurashige2009high}
Kurashige,~Y.; Yanai,~T. High-performance ab initio density matrix
  renormalization group method: Applicability to large-scale multireference
  problems for metal compounds. \emph{The Journal of chemical physics}
  \textbf{2009}, \emph{130}, 234114\relax
\mciteBstWouldAddEndPuncttrue
\mciteSetBstMidEndSepPunct{\mcitedefaultmidpunct}
{\mcitedefaultendpunct}{\mcitedefaultseppunct}\relax
\EndOfBibitem
\bibitem[Kurashige \latin{et~al.}(2013)Kurashige, Chan, and
  Yanai]{kurashige2013entangled}
Kurashige,~Y.; Chan,~G. K.-L.; Yanai,~T. Entangled quantum electronic
  wavefunctions of the Mn 4 CaO 5 cluster in photosystem II. \emph{Nature
  chemistry} \textbf{2013}, \emph{5}, 660\relax
\mciteBstWouldAddEndPuncttrue
\mciteSetBstMidEndSepPunct{\mcitedefaultmidpunct}
{\mcitedefaultendpunct}{\mcitedefaultseppunct}\relax
\EndOfBibitem
\bibitem[Sharma \latin{et~al.}(2014)Sharma, Sivalingam, Neese, and
  Chan]{sharma2014low}
Sharma,~S.; Sivalingam,~K.; Neese,~F.; Chan,~G. K.-L. Low-energy spectrum of
  iron--sulfur clusters directly from many-particle quantum mechanics.
  \emph{Nature chemistry} \textbf{2014}, \emph{6}, 927\relax
\mciteBstWouldAddEndPuncttrue
\mciteSetBstMidEndSepPunct{\mcitedefaultmidpunct}
{\mcitedefaultendpunct}{\mcitedefaultseppunct}\relax
\EndOfBibitem
\bibitem[Olivares-Amaya \latin{et~al.}(2015)Olivares-Amaya, Hu, Nakatani,
  Sharma, Yang, and Chan]{olivares2015ab}
Olivares-Amaya,~R.; Hu,~W.; Nakatani,~N.; Sharma,~S.; Yang,~J.; Chan,~G. K.-L.
  The ab-initio density matrix renormalization group in practice. \emph{The
  Journal of chemical physics} \textbf{2015}, \emph{142}, 034102\relax
\mciteBstWouldAddEndPuncttrue
\mciteSetBstMidEndSepPunct{\mcitedefaultmidpunct}
{\mcitedefaultendpunct}{\mcitedefaultseppunct}\relax
\EndOfBibitem
\bibitem[Mussard and Sharma(2017)Mussard, and Sharma]{mussard2017one}
Mussard,~B.; Sharma,~S. One-Step Treatment of Spin--Orbit Coupling and Electron
  Correlation in Large Active Spaces. \emph{Journal of chemical theory and
  computation} \textbf{2017}, \emph{14}, 154--165\relax
\mciteBstWouldAddEndPuncttrue
\mciteSetBstMidEndSepPunct{\mcitedefaultmidpunct}
{\mcitedefaultendpunct}{\mcitedefaultseppunct}\relax
\EndOfBibitem
\bibitem[Li \latin{et~al.}(2018)Li, Otten, Holmes, Sharma, and
  Umrigar]{Junhao2018}
Li,~J.; Otten,~M.; Holmes,~A.~A.; Sharma,~S.; Umrigar,~C.~J. Fast
  semistochastic heat-bath configuration interaction. \emph{J. Chem. Phys.}
  \textbf{2018}, \emph{149}, 214110\relax
\mciteBstWouldAddEndPuncttrue
\mciteSetBstMidEndSepPunct{\mcitedefaultmidpunct}
{\mcitedefaultendpunct}{\mcitedefaultseppunct}\relax
\EndOfBibitem
\bibitem[Booth \latin{et~al.}(2013)Booth, Gr{\"u}neis, Kresse, and
  Alavi]{booth2013towards}
Booth,~G.~H.; Gr{\"u}neis,~A.; Kresse,~G.; Alavi,~A. Towards an exact
  description of electronic wavefunctions in real solids. \emph{Nature}
  \textbf{2013}, \emph{493}, 365\relax
\mciteBstWouldAddEndPuncttrue
\mciteSetBstMidEndSepPunct{\mcitedefaultmidpunct}
{\mcitedefaultendpunct}{\mcitedefaultseppunct}\relax
\EndOfBibitem
\bibitem[Li~Manni and Alavi(2018)Li~Manni, and Alavi]{li2018understanding}
Li~Manni,~G.; Alavi,~A. Understanding the mechanism stabilizing intermediate
  spin states in Fe (II)-porphyrin. \emph{The Journal of Physical Chemistry A}
  \textbf{2018}, \emph{122}, 4935--4947\relax
\mciteBstWouldAddEndPuncttrue
\mciteSetBstMidEndSepPunct{\mcitedefaultmidpunct}
{\mcitedefaultendpunct}{\mcitedefaultseppunct}\relax
\EndOfBibitem
\bibitem[Andersson \latin{et~al.}(1990)Andersson, Malmqvist, Roos, Sadlej, and
  Wolinski]{andersson1990second}
Andersson,~K.; Malmqvist,~P.~A.; Roos,~B.~O.; Sadlej,~A.~J.; Wolinski,~K.
  Second-order perturbation theory with a CASSCF reference function.
  \emph{Journal of Physical Chemistry} \textbf{1990}, \emph{94},
  5483--5488\relax
\mciteBstWouldAddEndPuncttrue
\mciteSetBstMidEndSepPunct{\mcitedefaultmidpunct}
{\mcitedefaultendpunct}{\mcitedefaultseppunct}\relax
\EndOfBibitem
\bibitem[Andersson \latin{et~al.}(1992)Andersson, Malmqvist, and
  Roos]{andersson1992second}
Andersson,~K.; Malmqvist,~P.-{\AA}.; Roos,~B.~O. Second-order perturbation
  theory with a complete active space self-consistent field reference function.
  \emph{The Journal of chemical physics} \textbf{1992}, \emph{96},
  1218--1226\relax
\mciteBstWouldAddEndPuncttrue
\mciteSetBstMidEndSepPunct{\mcitedefaultmidpunct}
{\mcitedefaultendpunct}{\mcitedefaultseppunct}\relax
\EndOfBibitem
\bibitem[Angeli \latin{et~al.}(2001)Angeli, Cimiraglia, Evangelisti, Leininger,
  and Malrieu]{angeli2001introduction}
Angeli,~C.; Cimiraglia,~R.; Evangelisti,~S.; Leininger,~T.; Malrieu,~J.-P.
  Introduction of n-electron valence states for multireference perturbation
  theory. \emph{The Journal of Chemical Physics} \textbf{2001}, \emph{114},
  10252--10264\relax
\mciteBstWouldAddEndPuncttrue
\mciteSetBstMidEndSepPunct{\mcitedefaultmidpunct}
{\mcitedefaultendpunct}{\mcitedefaultseppunct}\relax
\EndOfBibitem
\bibitem[Angeli \latin{et~al.}(2002)Angeli, Cimiraglia, and
  Malrieu]{angeli2002n}
Angeli,~C.; Cimiraglia,~R.; Malrieu,~J.-P. n-electron valence state
  perturbation theory: A spinless formulation and an efficient implementation
  of the strongly contracted and of the partially contracted variants.
  \emph{The Journal of chemical physics} \textbf{2002}, \emph{117},
  9138--9153\relax
\mciteBstWouldAddEndPuncttrue
\mciteSetBstMidEndSepPunct{\mcitedefaultmidpunct}
{\mcitedefaultendpunct}{\mcitedefaultseppunct}\relax
\EndOfBibitem
\bibitem[Angeli \latin{et~al.}(2004)Angeli, Borini, Cestari, and
  Cimiraglia]{angeli2004quasidegenerate}
Angeli,~C.; Borini,~S.; Cestari,~M.; Cimiraglia,~R. A quasidegenerate
  formulation of the second order n-electron valence state perturbation theory
  approach. \emph{The Journal of chemical physics} \textbf{2004}, \emph{121},
  4043--4049\relax
\mciteBstWouldAddEndPuncttrue
\mciteSetBstMidEndSepPunct{\mcitedefaultmidpunct}
{\mcitedefaultendpunct}{\mcitedefaultseppunct}\relax
\EndOfBibitem
\bibitem[Werner and Knowles(1988)Werner, and Knowles]{werner1988efficient}
Werner,~H.-J.; Knowles,~P.~J. An efficient internally contracted
  multiconfiguration--reference configuration interaction method. \emph{The
  Journal of chemical physics} \textbf{1988}, \emph{89}, 5803--5814\relax
\mciteBstWouldAddEndPuncttrue
\mciteSetBstMidEndSepPunct{\mcitedefaultmidpunct}
{\mcitedefaultendpunct}{\mcitedefaultseppunct}\relax
\EndOfBibitem
\bibitem[Knowles and Werner(1988)Knowles, and Werner]{knowles1988efficient}
Knowles,~P.~J.; Werner,~H.-J. An efficient method for the evaluation of
  coupling coefficients in configuration interaction calculations.
  \emph{Chemical physics letters} \textbf{1988}, \emph{145}, 514--522\relax
\mciteBstWouldAddEndPuncttrue
\mciteSetBstMidEndSepPunct{\mcitedefaultmidpunct}
{\mcitedefaultendpunct}{\mcitedefaultseppunct}\relax
\EndOfBibitem
\bibitem[Knowles and Werner(1992)Knowles, and Werner]{knowles1992internally}
Knowles,~P.~J.; Werner,~H.-J. Internally contracted
  multiconfiguration-reference configuration interaction calculations for
  excited states. \emph{Theoretica chimica acta} \textbf{1992}, \emph{84},
  95--103\relax
\mciteBstWouldAddEndPuncttrue
\mciteSetBstMidEndSepPunct{\mcitedefaultmidpunct}
{\mcitedefaultendpunct}{\mcitedefaultseppunct}\relax
\EndOfBibitem
\bibitem[Bartlett and Musia{\l}(2007)Bartlett, and
  Musia{\l}]{bartlett2007coupled}
Bartlett,~R.~J.; Musia{\l},~M. Coupled-cluster theory in quantum chemistry.
  \emph{Reviews of Modern Physics} \textbf{2007}, \emph{79}, 291\relax
\mciteBstWouldAddEndPuncttrue
\mciteSetBstMidEndSepPunct{\mcitedefaultmidpunct}
{\mcitedefaultendpunct}{\mcitedefaultseppunct}\relax
\EndOfBibitem
\bibitem[Yanai and Chan(2006)Yanai, and Chan]{yanai2006canonical}
Yanai,~T.; Chan,~G. K.-L. Canonical transformation theory for multireference
  problems. \emph{The Journal of chemical physics} \textbf{2006}, \emph{124},
  194106\relax
\mciteBstWouldAddEndPuncttrue
\mciteSetBstMidEndSepPunct{\mcitedefaultmidpunct}
{\mcitedefaultendpunct}{\mcitedefaultseppunct}\relax
\EndOfBibitem
\bibitem[Neuscamman \latin{et~al.}(2010)Neuscamman, Yanai, and
  Chan]{neuscamman2010review}
Neuscamman,~E.; Yanai,~T.; Chan,~G. K.-L. A review of canonical transformation
  theory. \emph{International Reviews in Physical Chemistry} \textbf{2010},
  \emph{29}, 231--271\relax
\mciteBstWouldAddEndPuncttrue
\mciteSetBstMidEndSepPunct{\mcitedefaultmidpunct}
{\mcitedefaultendpunct}{\mcitedefaultseppunct}\relax
\EndOfBibitem
\bibitem[Evangelista(2014)]{evangelista2014driven}
Evangelista,~F.~A. A driven similarity renormalization group approach to
  quantum many-body problems. \emph{The Journal of chemical physics}
  \textbf{2014}, \emph{141}, 054109\relax
\mciteBstWouldAddEndPuncttrue
\mciteSetBstMidEndSepPunct{\mcitedefaultmidpunct}
{\mcitedefaultendpunct}{\mcitedefaultseppunct}\relax
\EndOfBibitem
\bibitem[Kurashige and Yanai(2011)Kurashige, and Yanai]{kurashige2011second}
Kurashige,~Y.; Yanai,~T. Second-order perturbation theory with a density matrix
  renormalization group self-consistent field reference function: Theory and
  application to the study of chromium dimer. \emph{The Journal of chemical
  physics} \textbf{2011}, \emph{135}, 094104\relax
\mciteBstWouldAddEndPuncttrue
\mciteSetBstMidEndSepPunct{\mcitedefaultmidpunct}
{\mcitedefaultendpunct}{\mcitedefaultseppunct}\relax
\EndOfBibitem
\bibitem[Guo \latin{et~al.}(2016)Guo, Watson, Hu, Sun, and Chan]{guo2016n}
Guo,~S.; Watson,~M.~A.; Hu,~W.; Sun,~Q.; Chan,~G. K.-L. N-electron valence
  state perturbation theory based on a density matrix renormalization group
  reference function, with applications to the chromium dimer and a trimer
  model of poly (p-phenylenevinylene). \emph{Journal of chemical theory and
  computation} \textbf{2016}, \emph{12}, 1583--1591\relax
\mciteBstWouldAddEndPuncttrue
\mciteSetBstMidEndSepPunct{\mcitedefaultmidpunct}
{\mcitedefaultendpunct}{\mcitedefaultseppunct}\relax
\EndOfBibitem
\bibitem[Roemelt \latin{et~al.}(2016)Roemelt, Guo, and
  Chan]{roemelt2016projected}
Roemelt,~M.; Guo,~S.; Chan,~G. K.-L. A projected approximation to strongly
  contracted N-electron valence perturbation theory for DMRG wavefunctions.
  \emph{The Journal of chemical physics} \textbf{2016}, \emph{144},
  204113\relax
\mciteBstWouldAddEndPuncttrue
\mciteSetBstMidEndSepPunct{\mcitedefaultmidpunct}
{\mcitedefaultendpunct}{\mcitedefaultseppunct}\relax
\EndOfBibitem
\bibitem[Zgid \latin{et~al.}(2009)Zgid, Ghosh, Neuscamman, and
  Chan]{zgid2009study}
Zgid,~D.; Ghosh,~D.; Neuscamman,~E.; Chan,~G. K.-L. A study of cumulant
  approximations to n-electron valence multireference perturbation theory.
  \emph{The Journal of chemical physics} \textbf{2009}, \emph{130},
  194107\relax
\mciteBstWouldAddEndPuncttrue
\mciteSetBstMidEndSepPunct{\mcitedefaultmidpunct}
{\mcitedefaultendpunct}{\mcitedefaultseppunct}\relax
\EndOfBibitem
\bibitem[Saitow \latin{et~al.}(2013)Saitow, Kurashige, and
  Yanai]{saitow2013multireference}
Saitow,~M.; Kurashige,~Y.; Yanai,~T. Multireference configuration interaction
  theory using cumulant reconstruction with internal contraction of density
  matrix renormalization group wave function. \emph{The Journal of chemical
  physics} \textbf{2013}, \emph{139}, 044118\relax
\mciteBstWouldAddEndPuncttrue
\mciteSetBstMidEndSepPunct{\mcitedefaultmidpunct}
{\mcitedefaultendpunct}{\mcitedefaultseppunct}\relax
\EndOfBibitem
\bibitem[Kurashige \latin{et~al.}(2014)Kurashige, Chalupsk{\`y}, Lan, and
  Yanai]{kurashige2014complete}
Kurashige,~Y.; Chalupsk{\`y},~J.; Lan,~T.~N.; Yanai,~T. Complete active space
  second-order perturbation theory with cumulant approximation for extended
  active-space wavefunction from density matrix renormalization group.
  \emph{The Journal of chemical physics} \textbf{2014}, \emph{141},
  174111\relax
\mciteBstWouldAddEndPuncttrue
\mciteSetBstMidEndSepPunct{\mcitedefaultmidpunct}
{\mcitedefaultendpunct}{\mcitedefaultseppunct}\relax
\EndOfBibitem
\bibitem[Saitow \latin{et~al.}(2015)Saitow, Kurashige, and
  Yanai]{saitow2015fully}
Saitow,~M.; Kurashige,~Y.; Yanai,~T. Fully internally contracted multireference
  configuration interaction theory using density matrix renormalization group:
  A reduced-scaling implementation derived by computer-aided tensor
  factorization. \emph{Journal of chemical theory and computation}
  \textbf{2015}, \emph{11}, 5120--5131\relax
\mciteBstWouldAddEndPuncttrue
\mciteSetBstMidEndSepPunct{\mcitedefaultmidpunct}
{\mcitedefaultendpunct}{\mcitedefaultseppunct}\relax
\EndOfBibitem
\bibitem[Shirai \latin{et~al.}(2016)Shirai, Kurashige, and
  Yanai]{shirai2016computational}
Shirai,~S.; Kurashige,~Y.; Yanai,~T. Computational evidence of inversion of 1La
  and 1Lb-derived excited states in naphthalene excimer formation from ab
  Initio multireference theory with large active space: DMRG-CASPT2 Study.
  \emph{Journal of chemical theory and computation} \textbf{2016}, \emph{12},
  2366--2372\relax
\mciteBstWouldAddEndPuncttrue
\mciteSetBstMidEndSepPunct{\mcitedefaultmidpunct}
{\mcitedefaultendpunct}{\mcitedefaultseppunct}\relax
\EndOfBibitem
\bibitem[Phung \latin{et~al.}(2016)Phung, Wouters, and
  Pierloot]{phung2016cumulant}
Phung,~Q.~M.; Wouters,~S.; Pierloot,~K. Cumulant approximated second-Order
  perturbation theory based on the density matrix renormalization group for
  transition metal complexes: a benchmark study. \emph{Journal of chemical
  theory and computation} \textbf{2016}, \emph{12}, 4352--4361\relax
\mciteBstWouldAddEndPuncttrue
\mciteSetBstMidEndSepPunct{\mcitedefaultmidpunct}
{\mcitedefaultendpunct}{\mcitedefaultseppunct}\relax
\EndOfBibitem
\bibitem[Yanai \latin{et~al.}(2017)Yanai, Saitow, Xiong, Chalupský, Kurashige,
  Guo, and Sharma]{yanai2017multistate}
Yanai,~T.; Saitow,~M.; Xiong,~X.-G.; Chalupský,~J.; Kurashige,~Y.; Guo,~S.;
  Sharma,~S. Multistate Complete-Active-Space Second-Order Perturbation Theory
  Based on Density Matrix Renormalization Group Reference States. \emph{Journal
  of chemical theory and computation} \textbf{2017}, \emph{13},
  4829--4840\relax
\mciteBstWouldAddEndPuncttrue
\mciteSetBstMidEndSepPunct{\mcitedefaultmidpunct}
{\mcitedefaultendpunct}{\mcitedefaultseppunct}\relax
\EndOfBibitem
\bibitem[Nakatani and Guo(2017)Nakatani, and Guo]{nakatani-caspt2}
Nakatani,~N.; Guo,~S. Density matrix renormalization group (DMRG) method as a
  common tool for large active-space CASSCF/CASPT2 calculations. \emph{The
  Journal of Chemical Physics} \textbf{2017}, \emph{146}, 094102\relax
\mciteBstWouldAddEndPuncttrue
\mciteSetBstMidEndSepPunct{\mcitedefaultmidpunct}
{\mcitedefaultendpunct}{\mcitedefaultseppunct}\relax
\EndOfBibitem
\bibitem[Wouters \latin{et~al.}(2016)Wouters, Van~Speybroeck, and
  Van~Neck]{wouters2016dmrg}
Wouters,~S.; Van~Speybroeck,~V.; Van~Neck,~D. DMRG-CASPT2 study of the
  longitudinal static second hyperpolarizability of all-trans polyenes.
  \emph{The Journal of chemical physics} \textbf{2016}, \emph{145},
  054120\relax
\mciteBstWouldAddEndPuncttrue
\mciteSetBstMidEndSepPunct{\mcitedefaultmidpunct}
{\mcitedefaultendpunct}{\mcitedefaultseppunct}\relax
\EndOfBibitem
\bibitem[Celani and Werner(2000)Celani, and Werner]{celani2000multireference}
Celani,~P.; Werner,~H.-J. Multireference perturbation theory for large
  restricted and selected active space reference wave functions. \emph{J. Chem.
  Phys.} \textbf{2000}, \emph{112}, 5546--5557\relax
\mciteBstWouldAddEndPuncttrue
\mciteSetBstMidEndSepPunct{\mcitedefaultmidpunct}
{\mcitedefaultendpunct}{\mcitedefaultseppunct}\relax
\EndOfBibitem
\bibitem[Shamasundar \latin{et~al.}(2011)Shamasundar, Knizia, and
  Werner]{shamasundar2011new}
Shamasundar,~K.; Knizia,~G.; Werner,~H.-J. A new internally contracted
  multi-reference configuration interaction method. \emph{The Journal of
  chemical physics} \textbf{2011}, \emph{135}, 054101\relax
\mciteBstWouldAddEndPuncttrue
\mciteSetBstMidEndSepPunct{\mcitedefaultmidpunct}
{\mcitedefaultendpunct}{\mcitedefaultseppunct}\relax
\EndOfBibitem
\bibitem[Sharma and Chan(2014)Sharma, and Chan]{sharma2014communication}
Sharma,~S.; Chan,~G. K.-L. Communication: A flexible multi-reference
  perturbation theory by minimizing the Hylleraas functional with matrix
  product states. 2014\relax
\mciteBstWouldAddEndPuncttrue
\mciteSetBstMidEndSepPunct{\mcitedefaultmidpunct}
{\mcitedefaultendpunct}{\mcitedefaultseppunct}\relax
\EndOfBibitem
\bibitem[Sharma \latin{et~al.}(2017)Sharma, Knizia, Guo, and
  Alavi]{sharma2017combining}
Sharma,~S.; Knizia,~G.; Guo,~S.; Alavi,~A. Combining internally contracted
  states and matrix product states to perform multireference perturbation
  theory. \emph{Journal of chemical theory and computation} \textbf{2017},
  \emph{13}, 488--498\relax
\mciteBstWouldAddEndPuncttrue
\mciteSetBstMidEndSepPunct{\mcitedefaultmidpunct}
{\mcitedefaultendpunct}{\mcitedefaultseppunct}\relax
\EndOfBibitem
\bibitem[Sokolov \latin{et~al.}(2017)Sokolov, Guo, Ronca, and
  Chan]{sokolov2017time}
Sokolov,~A.~Y.; Guo,~S.; Ronca,~E.; Chan,~G. K.-L. Time-dependent N-electron
  valence perturbation theory with matrix product state reference wavefunctions
  for large active spaces and basis sets: Applications to the chromium dimer
  and all-trans polyenes. \emph{The Journal of chemical physics} \textbf{2017},
  \emph{146}, 244102\relax
\mciteBstWouldAddEndPuncttrue
\mciteSetBstMidEndSepPunct{\mcitedefaultmidpunct}
{\mcitedefaultendpunct}{\mcitedefaultseppunct}\relax
\EndOfBibitem
\bibitem[Sharma and Alavi(2015)Sharma, and Alavi]{sharma2015multireference}
Sharma,~S.; Alavi,~A. Multireference linearized coupled cluster theory for
  strongly correlated systems using matrix product states. \emph{The Journal of
  chemical physics} \textbf{2015}, \emph{143}, 102815\relax
\mciteBstWouldAddEndPuncttrue
\mciteSetBstMidEndSepPunct{\mcitedefaultmidpunct}
{\mcitedefaultendpunct}{\mcitedefaultseppunct}\relax
\EndOfBibitem
\bibitem[Sharma \latin{et~al.}(2016)Sharma, Jeanmairet, and
  Alavi]{sharma2016quasi}
Sharma,~S.; Jeanmairet,~G.; Alavi,~A. Quasi-degenerate perturbation theory
  using matrix product states. \emph{The Journal of chemical physics}
  \textbf{2016}, \emph{144}, 034103\relax
\mciteBstWouldAddEndPuncttrue
\mciteSetBstMidEndSepPunct{\mcitedefaultmidpunct}
{\mcitedefaultendpunct}{\mcitedefaultseppunct}\relax
\EndOfBibitem
\bibitem[Tahara and Imada(2008)Tahara, and Imada]{tahara2008variational}
Tahara,~D.; Imada,~M. Variational Monte Carlo method combined with
  quantum-number projection and multi-variable optimization. \emph{J. Phys.
  Soc. Jpn.} \textbf{2008}, \emph{77}, 114701\relax
\mciteBstWouldAddEndPuncttrue
\mciteSetBstMidEndSepPunct{\mcitedefaultmidpunct}
{\mcitedefaultendpunct}{\mcitedefaultseppunct}\relax
\EndOfBibitem
\bibitem[Neuscamman(2012)]{neuscamman2012size}
Neuscamman,~E. Size consistency error in the antisymmetric geminal power wave
  function can be completely removed. \emph{Phys. Rev. Lett.} \textbf{2012},
  \emph{109}, 203001\relax
\mciteBstWouldAddEndPuncttrue
\mciteSetBstMidEndSepPunct{\mcitedefaultmidpunct}
{\mcitedefaultendpunct}{\mcitedefaultseppunct}\relax
\EndOfBibitem
\bibitem[Neuscamman(2013)]{neuscamman2013jastrow}
Neuscamman,~E. The Jastrow antisymmetric geminal power in Hilbert space:
  Theory, benchmarking, and application to a novel transition state. \emph{J.
  Chem. Phys.} \textbf{2013}, \emph{139}, 194105\relax
\mciteBstWouldAddEndPuncttrue
\mciteSetBstMidEndSepPunct{\mcitedefaultmidpunct}
{\mcitedefaultendpunct}{\mcitedefaultseppunct}\relax
\EndOfBibitem
\bibitem[Mahajan and Sharma(2019)Mahajan, and Sharma]{mahajan2019symmetry}
Mahajan,~A.; Sharma,~S. Symmetry-Projected Jastrow Mean-Field Wave Function in
  Variational Monte Carlo. \emph{The Journal of Physical Chemistry A}
  \textbf{2019}, \emph{123}, 3911--3921\relax
\mciteBstWouldAddEndPuncttrue
\mciteSetBstMidEndSepPunct{\mcitedefaultmidpunct}
{\mcitedefaultendpunct}{\mcitedefaultseppunct}\relax
\EndOfBibitem
\bibitem[McLean and Liu(1973)McLean, and Liu]{mclean1973classification}
McLean,~A.; Liu,~B. Classification of configurations and the determination of
  interacting and noninteracting spaces in configuration interaction. \emph{The
  Journal of Chemical Physics} \textbf{1973}, \emph{58}, 1066--1078\relax
\mciteBstWouldAddEndPuncttrue
\mciteSetBstMidEndSepPunct{\mcitedefaultmidpunct}
{\mcitedefaultendpunct}{\mcitedefaultseppunct}\relax
\EndOfBibitem
\bibitem[Meyer(1977)]{Meyer1977}
Meyer,~W. \emph{{Modern Theoretical Chemistry}}; Plenum Press New York,
  1977\relax
\mciteBstWouldAddEndPuncttrue
\mciteSetBstMidEndSepPunct{\mcitedefaultmidpunct}
{\mcitedefaultendpunct}{\mcitedefaultseppunct}\relax
\EndOfBibitem
\bibitem[Siegbahn(1980)]{siegbahn1980direct}
Siegbahn,~P.~E. Direct configuration interaction with a reference state
  composed of many reference configurations. \emph{International Journal of
  Quantum Chemistry} \textbf{1980}, \emph{18}, 1229--1242\relax
\mciteBstWouldAddEndPuncttrue
\mciteSetBstMidEndSepPunct{\mcitedefaultmidpunct}
{\mcitedefaultendpunct}{\mcitedefaultseppunct}\relax
\EndOfBibitem
\bibitem[Sivalingam \latin{et~al.}(2016)Sivalingam, Krupicka, Auer, and
  Neese]{sivalingam2016comparison}
Sivalingam,~K.; Krupicka,~M.; Auer,~A.~A.; Neese,~F. Comparison of fully
  internally and strongly contracted multireference configuration interaction
  procedures. \emph{The Journal of chemical physics} \textbf{2016}, \emph{145},
  054104\relax
\mciteBstWouldAddEndPuncttrue
\mciteSetBstMidEndSepPunct{\mcitedefaultmidpunct}
{\mcitedefaultendpunct}{\mcitedefaultseppunct}\relax
\EndOfBibitem
\bibitem[Luo \latin{et~al.}(2018)Luo, Ma, Wang, and Ma]{haibo}
Luo,~Z.; Ma,~Y.; Wang,~X.; Ma,~H. Externally-Contracted Multireference
  Configuration Interaction Method Using a DMRG Reference Wave Function.
  \emph{Journal of Chemical Theory and Computation} \textbf{2018}, \emph{14},
  4747--4755\relax
\mciteBstWouldAddEndPuncttrue
\mciteSetBstMidEndSepPunct{\mcitedefaultmidpunct}
{\mcitedefaultendpunct}{\mcitedefaultseppunct}\relax
\EndOfBibitem
\bibitem[Bortz \latin{et~al.}(1975)Bortz, Kalos, and Lebowitz]{BORTZ197510}
Bortz,~A.; Kalos,~M.; Lebowitz,~J. A new algorithm for Monte Carlo simulation
  of Ising spin systems. \emph{J. Comput. Phys.} \textbf{1975}, \emph{17}, 10
  -- 18\relax
\mciteBstWouldAddEndPuncttrue
\mciteSetBstMidEndSepPunct{\mcitedefaultmidpunct}
{\mcitedefaultendpunct}{\mcitedefaultseppunct}\relax
\EndOfBibitem
\bibitem[Gillespie(1976)]{GILLESPIE1976403}
Gillespie,~D.~T. A general method for numerically simulating the stochastic
  time evolution of coupled chemical reactions. \emph{J. Comp. Phys.}
  \textbf{1976}, \emph{22}, 403 -- 434\relax
\mciteBstWouldAddEndPuncttrue
\mciteSetBstMidEndSepPunct{\mcitedefaultmidpunct}
{\mcitedefaultendpunct}{\mcitedefaultseppunct}\relax
\EndOfBibitem
\bibitem[Sabzevari and Sharma(2018)Sabzevari, and Sharma]{Sabzevari18}
Sabzevari,~I.; Sharma,~S. Improved Speed and Scaling in Orbital Space
  Variational Monte Carlo. \emph{J. Chem. Theory Comput.} \textbf{2018},
  \emph{14}, 6276--6286\relax
\mciteBstWouldAddEndPuncttrue
\mciteSetBstMidEndSepPunct{\mcitedefaultmidpunct}
{\mcitedefaultendpunct}{\mcitedefaultseppunct}\relax
\EndOfBibitem
\bibitem[Nightingale and Melik-Alaverdian(2001)Nightingale, and
  Melik-Alaverdian]{nightingale2001optimization}
Nightingale,~M.; Melik-Alaverdian,~V. Optimization of ground-and excited-state
  wave functions and van der Waals clusters. \emph{Physical review letters}
  \textbf{2001}, \emph{87}, 043401\relax
\mciteBstWouldAddEndPuncttrue
\mciteSetBstMidEndSepPunct{\mcitedefaultmidpunct}
{\mcitedefaultendpunct}{\mcitedefaultseppunct}\relax
\EndOfBibitem
\bibitem[Umrigar \latin{et~al.}(2007)Umrigar, Toulouse, Filippi, Sorella, and
  Hennig]{umrigar2007alleviation}
Umrigar,~C.; Toulouse,~J.; Filippi,~C.; Sorella,~S.; Hennig,~R.~G. Alleviation
  of the fermion-sign problem by optimization of many-body wave functions.
  \emph{Physical review letters} \textbf{2007}, \emph{98}, 110201\relax
\mciteBstWouldAddEndPuncttrue
\mciteSetBstMidEndSepPunct{\mcitedefaultmidpunct}
{\mcitedefaultendpunct}{\mcitedefaultseppunct}\relax
\EndOfBibitem
\bibitem[Toulouse and Umrigar(2007)Toulouse, and
  Umrigar]{toulouse2007optimization}
Toulouse,~J.; Umrigar,~C.~J. Optimization of quantum Monte Carlo wave functions
  by energy minimization. \emph{The Journal of chemical physics} \textbf{2007},
  \emph{126}, 084102\relax
\mciteBstWouldAddEndPuncttrue
\mciteSetBstMidEndSepPunct{\mcitedefaultmidpunct}
{\mcitedefaultendpunct}{\mcitedefaultseppunct}\relax
\EndOfBibitem
\bibitem[Toulouse and Umrigar(2008)Toulouse, and Umrigar]{toulouse2008full}
Toulouse,~J.; Umrigar,~C. Full optimization of Jastrow--Slater wave functions
  with application to the first-row atoms and homonuclear diatomic molecules.
  \emph{The Journal of chemical physics} \textbf{2008}, \emph{128},
  174101\relax
\mciteBstWouldAddEndPuncttrue
\mciteSetBstMidEndSepPunct{\mcitedefaultmidpunct}
{\mcitedefaultendpunct}{\mcitedefaultseppunct}\relax
\EndOfBibitem
\bibitem[Zhao and Neuscamman(2017)Zhao, and Neuscamman]{zhao2017blocked}
Zhao,~L.; Neuscamman,~E. A blocked linear method for optimizing large parameter
  sets in variational monte carlo. \emph{Journal of chemical theory and
  computation} \textbf{2017}, \emph{13}, 2604--2611\relax
\mciteBstWouldAddEndPuncttrue
\mciteSetBstMidEndSepPunct{\mcitedefaultmidpunct}
{\mcitedefaultendpunct}{\mcitedefaultseppunct}\relax
\EndOfBibitem
\bibitem[Sabzevari \latin{et~al.}(2019)Sabzevari, Mahajan, and
  Sharma]{sabzevari2019accelerated}
Sabzevari,~I.; Mahajan,~A.; Sharma,~S. An accelerated linear method for
  optimizing non-linear wavefunctions in variational Monte Carlo. \emph{arXiv
  preprint arXiv:1908.04423} \textbf{2019}, \relax
\mciteBstWouldAddEndPunctfalse
\mciteSetBstMidEndSepPunct{\mcitedefaultmidpunct}
{}{\mcitedefaultseppunct}\relax
\EndOfBibitem
\bibitem[Reddi \latin{et~al.}(2018)Reddi, Kale, and Kumar]{Reddi2018}
Reddi,~S.~J.; Kale,~S.; Kumar,~S. On the Convergence of Adam and Beyond.
  International Conference on Learning Representations. 2018; pp 1--23\relax
\mciteBstWouldAddEndPuncttrue
\mciteSetBstMidEndSepPunct{\mcitedefaultmidpunct}
{\mcitedefaultendpunct}{\mcitedefaultseppunct}\relax
\EndOfBibitem
\bibitem[Schwarz \latin{et~al.}(2017)Schwarz, Alavi, and
  Booth]{schwarz2017projector}
Schwarz,~L.~R.; Alavi,~A.; Booth,~G.~H. Projector Quantum Monte Carlo Method
  for Nonlinear Wave Functions. \emph{Physical review letters} \textbf{2017},
  \emph{118}, 176403\relax
\mciteBstWouldAddEndPuncttrue
\mciteSetBstMidEndSepPunct{\mcitedefaultmidpunct}
{\mcitedefaultendpunct}{\mcitedefaultseppunct}\relax
\EndOfBibitem
\bibitem[Otis and Neuscamman(2019)Otis, and Neuscamman]{otis2019complementary}
Otis,~L.; Neuscamman,~E. Complementary First and Second Derivative Methods for
  Ansatz Optimization in Variational Monte Carlo. \emph{Physical Chemistry
  Chemical Physics} \textbf{2019}, \relax
\mciteBstWouldAddEndPunctfalse
\mciteSetBstMidEndSepPunct{\mcitedefaultmidpunct}
{}{\mcitedefaultseppunct}\relax
\EndOfBibitem
\bibitem[Dyall(1995)]{dyall1995choice}
Dyall,~K.~G. The choice of a zeroth-order Hamiltonian for second-order
  perturbation theory with a complete active space self-consistent-field
  reference function. \emph{The Journal of chemical physics} \textbf{1995},
  \emph{102}, 4909--4918\relax
\mciteBstWouldAddEndPuncttrue
\mciteSetBstMidEndSepPunct{\mcitedefaultmidpunct}
{\mcitedefaultendpunct}{\mcitedefaultseppunct}\relax
\EndOfBibitem
\bibitem[Werner \latin{et~al.}(2012)Werner, Knowles, Knizia, Manby, and
  Sch{\"u}tz]{werner2012molpro}
Werner,~H.-J.; Knowles,~P.~J.; Knizia,~G.; Manby,~F.~R.; Sch{\"u}tz,~M. Molpro:
  a general-purpose quantum chemistry program package. \emph{WIREs Comput. Mol.
  Sci.} \textbf{2012}, \emph{2}, 242--253\relax
\mciteBstWouldAddEndPuncttrue
\mciteSetBstMidEndSepPunct{\mcitedefaultmidpunct}
{\mcitedefaultendpunct}{\mcitedefaultseppunct}\relax
\EndOfBibitem
\bibitem[Sun \latin{et~al.}(2018)Sun, Berkelbach, Blunt, Booth, Guo, Li, Liu,
  McClain, Sayfutyarova, Sharma, Wouters, and Chan]{sun2018pyscf}
Sun,~Q.; Berkelbach,~T.~C.; Blunt,~N.~S.; Booth,~G.~H.; Guo,~S.; Li,~Z.;
  Liu,~J.; McClain,~J.~D.; Sayfutyarova,~E.~R.; Sharma,~S.; Wouters,~S.;
  Chan,~K.-L.~G. PySCF: the Python-based simulations of chemistry framework.
  \emph{WIREs Comput. Mol. Sci.} \textbf{2018}, \emph{8}, e1340\relax
\mciteBstWouldAddEndPuncttrue
\mciteSetBstMidEndSepPunct{\mcitedefaultmidpunct}
{\mcitedefaultendpunct}{\mcitedefaultseppunct}\relax
\EndOfBibitem
\bibitem[Zhao and Neuscamman(2016)Zhao, and Neuscamman]{Zhao2016}
Zhao,~L.; Neuscamman,~E. Equation of Motion Theory for Excited States in
  Variational Monte Carlo and the Jastrow Antisymmetric Geminal Power in
  Hilbert Space. \emph{Journal of Chemical Theory and Computation}
  \textbf{2016}, \emph{12}, 3719--3726\relax
\mciteBstWouldAddEndPuncttrue
\mciteSetBstMidEndSepPunct{\mcitedefaultmidpunct}
{\mcitedefaultendpunct}{\mcitedefaultseppunct}\relax
\EndOfBibitem
\bibitem[Blunt \latin{et~al.}(2018)Blunt, Alavi, and Booth]{blunt2018nonlinear}
Blunt,~N.~S.; Alavi,~A.; Booth,~G.~H. Nonlinear biases, stochastically sampled
  effective Hamiltonians, and spectral functions in quantum Monte Carlo
  methods. \emph{Physical Review B} \textbf{2018}, \emph{98}, 085118\relax
\mciteBstWouldAddEndPuncttrue
\mciteSetBstMidEndSepPunct{\mcitedefaultmidpunct}
{\mcitedefaultendpunct}{\mcitedefaultseppunct}\relax
\EndOfBibitem
\bibitem[Angeli \latin{et~al.}(2012)Angeli, Cimiraglia, and
  Pastore]{angeli2012comparison}
Angeli,~C.; Cimiraglia,~R.; Pastore,~M. A comparison of various approaches in
  internally contracted multireference configuration interaction: the carbon
  dimer as a test case. \emph{Molecular Physics} \textbf{2012}, \emph{110},
  2963--2968\relax
\mciteBstWouldAddEndPuncttrue
\mciteSetBstMidEndSepPunct{\mcitedefaultmidpunct}
{\mcitedefaultendpunct}{\mcitedefaultseppunct}\relax
\EndOfBibitem
\end{mcitethebibliography}

\providecommand{\latin}[1]{#1}
\makeatletter
\providecommand{\doi}
  {\begingroup\let\do\@makeother\dospecials
  \catcode`\{=1 \catcode`\}=2 \doi@aux}
\providecommand{\doi@aux}[1]{\endgroup\texttt{#1}}
\makeatother
\providecommand*\mcitethebibliography{\thebibliography}
\csname @ifundefined\endcsname{endmcitethebibliography}
  {\let\endmcitethebibliography\endthebibliography}{}

\end{document}